\documentclass[cameraready]{Interspeech}
\usepackage{cite}
\usepackage{amsmath}
\usepackage{multirow}
\usepackage{subcaption}

\title{INSPIRE: A Benchmark for Instruction-Aware Speech Retrieval}

\author[affiliation={1}]{Chen-An Li}{}
\author[affiliation={1,2}]{Hung-yi Lee}{}

\address{
  $^1$National Taiwan University \\
  $^2$NTU Artificial Intelligence Center of Research Excellence (NTU AI-CoRE)
}

\email{\{f13942069, hungyilee\}@ntu.edu.tw}

\keywords{speech retrieval, instruction-following}

\usepackage{comment}

\begin{document}

\maketitle

\begin{abstract}
Existing speech retrieval systems rely on fixed similarity matching and cannot adapt to diverse user intents. 
We introduce INSPIRE, the first benchmark for instruction-aware speech retrieval, in which natural-language instructions dynamically specify relevance criteria, including semantic content, speaker identity, speaking style, environmental sounds, and their combinations. 
We evaluate four retrieval paradigms: large audio-language models, cascaded pipelines, self-supervised speech models, and contrastive audio-language models. 
Our results reveal that no current method robustly handles all retrieval intents. 
Text-based approaches perform relatively better at semantic retrieval but struggle with paralinguistic attributes, while speech-based models are moderately better at capturing acoustic properties but falter at following instructions. 
These findings highlight the need for unified architectures capable of instruction-aware speech retrieval.
\end{abstract}

\section{Introduction}
\label{sec:introduction}

Speech retrieval, the task of searching a collection of spoken documents to find those relevant to a user's spoken query, is traditionally built around fixed notions of similarity, matching documents that are acoustically or semantically close to a query~\cite{lee2015spoken, lin2024speechdpr}. 
However, real-world retrieval needs are often instruction-driven. 
For instance, given a single reference audio clip, a customer service manager might want to find other calls exhibiting the same frustrated tone, a journalist may search for archival statements made by that exact speaker, or an investigator might look for clips recorded in a similarly noisy background. 
Each scenario defines relevance in entirely different ways, yet existing systems apply a single, rigid similarity function.
We therefore argue that speech retrieval should move beyond similarity-only paradigms toward an instruction-aware framework that explicitly specifies relevance criteria. 
Despite its success in text retrieval~\cite{asai2023task, jiang2024e5}, instruction-following remains largely unexplored for speech retrieval.

\begin{figure}[hbtp]
    \centering    
    \includegraphics[width=0.95\linewidth]{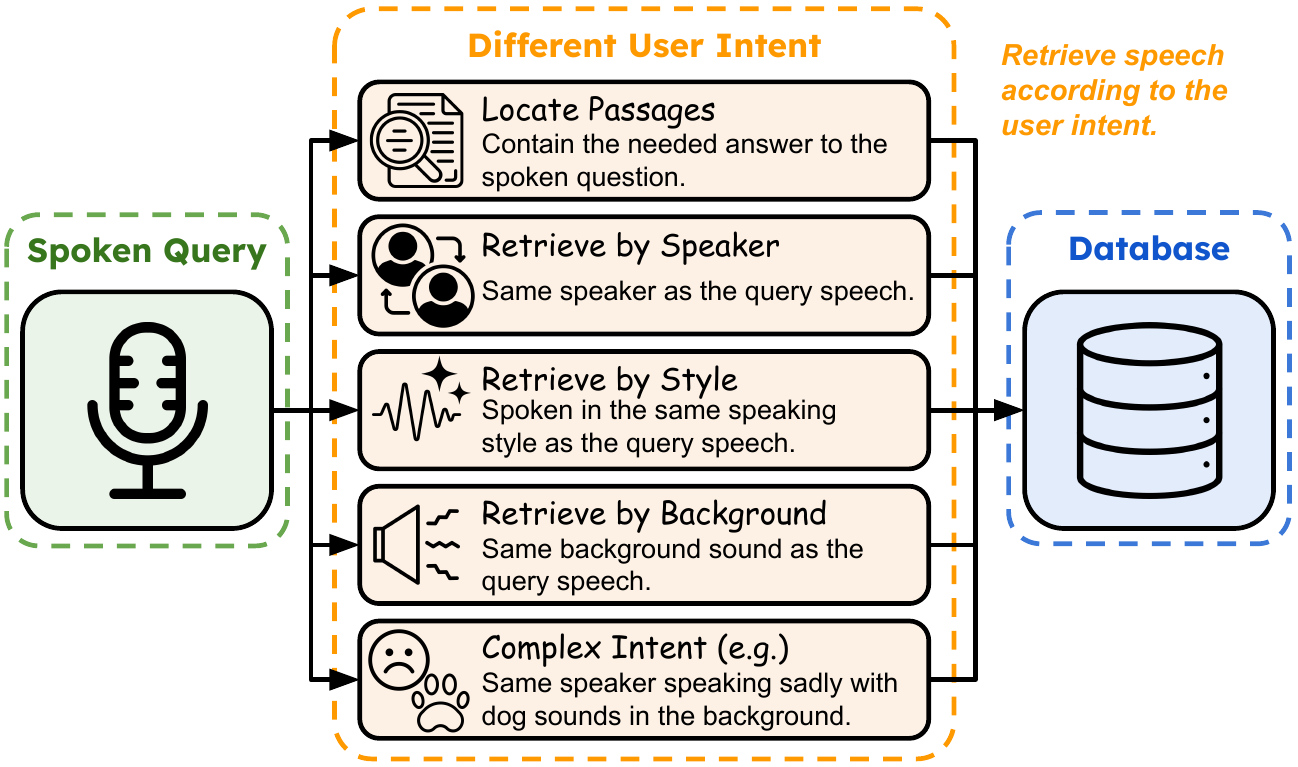}
    \vspace{-5pt}
    \caption{Illustrating instruction-aware speech retrieval across diverse user intents such as content, speaker, style, background, and their combinations.}
    \vspace{-8pt}
    \label{fig:overview}
\end{figure}

Compared to text retrieval, instruction-aware speech retrieval is challenging because the query is a speech signal and the relevant attributes are heterogeneous.
The instruction may refer to linguistic content (``what is said''), paralinguistic characteristics (``how it is said''), or environmental context (``where it is said''), and the system must weigh these cues differently depending on the instruction.
This demands representations that preserve fine-grained acoustic detail without sacrificing compositional instruction following.

As illustrated in Figure~\ref{fig:overview}, the same spoken query paired with different instructions should retrieve different target spoken documents.
The example in Figure~\ref{fig:overview} shows content-based retrieval, speaker-based retrieval, style-based retrieval, background-based retrieval, and composite intents that combine multiple attributes.

In this paper, we introduce INSPIRE\footnote{The code and data are publicly available at \url{https://github.com/lca0503/INSPIRE}.}, a benchmark for Instruction-Aware Speech Retrieval that systematically evaluates whether retrieval systems can follow natural-language instructions when selecting relevant spoken documents.
INSPIRE targets instruction-aware retrieval, where the instruction dynamically describes relevance.

Our primary contribution is formalizing instruction-aware speech retrieval as a benchmarked research problem.
Specifically, we:
\begin{itemize}
    \item Introduce INSPIRE, the first benchmark for instruction-aware speech retrieval spanning semantic, speaker, style, background, and multi-attribute intents;
    \item Provide a unified evaluation protocol that compares diverse model families under the same instruction formulation;
    \item Conduct an extensive empirical study covering large audio-language models, cascaded pipelines combining ASR and captioning with text retrieval, self-supervised speech representations, and contrastive audio-language architectures;
    \item Analyze failure modes across instruction types and subsets, revealing that no existing method robustly handles all aspects of instruction-aware speech retrieval within a single unified framework.
\end{itemize}

These findings highlight the need for new architectures and training paradigms specifically tailored to instruction-aware speech retrieval. 
We hope this work motivates future research to enable users to search speech collections with the same naturalness and precision as in text and image retrieval. 
\section{Related Work}
\label{sec:related_work}

\subsection{Spoken Content Retrieval}
Early work focused on query-by-example spoken term detection~\cite{ao2018query, ram2018cnn, ram2020neural}, where users provide a spoken sample to locate specific words or phrases in speech.
Subsequent research extended these matching-based methods to spoken content retrieval~\cite{lee2015spoken, lin2024speechdpr, witbrock1997speech, garofolo2000trec, chelba2008retrieval, larson2012spoken, hu25b_interspeech}, including speech-to-speech retrieval and text-to-speech retrieval over large speech collections.
These lines of work typically define relevance through acoustic or semantic similarity, with a fixed retrieval objective per spoken query.
In contrast, INSPIRE considers settings where an instruction defines relevance on-the-fly, potentially across multiple attributes, requiring the model to resolve which dimension of similarity cues takes priority.

\subsection{Spoken Question Answering}
Spoken question answering (SQA)~\cite{turmo2009overview, comas2012sibyl, li2018spoken, lee2018odsqa, lin22c_interspeech, you2022end, shon2023slue} identifies answer spans within spoken passages for a given question.
While SQA captures an important semantic retrieval use case, it does not address instruction-aware retrieval, in which relevance may depend on paralinguistic attributes.
INSPIRE includes SQA-like semantic intents as one subset but goes further, requiring models to handle speaker-, style-, and environment-based constraints alongside multi-attribute compositions, revealing whether models that excel at semantic retrieval generalize when relevance shifts beyond content.

\subsection{Instruction-Aware Retrieval in Other Modalities}
Instruction-aware retrieval has recently become a strong paradigm in text~\cite{asai2023task, su2023one, wang2024improving, lee2025nvembed, lee2025gemini, zhang2025qwen3} and image~\cite{jiang2024e5, wu2021fashion, liu2021image, baldrati2023zero, saito2023pic2word, zhang2024magiclens,  zhang2025bridging, zhou2025megapairs, jiang2025vlmvec, meng2025vlm2vec, li2026qwen3} retrieval.
In contrast, instruction-following has not been systematically studied for speech retrieval, motivating the INSPIRE benchmark.
INSPIRE adopts the same high-level principle, using a natural-language instruction to specify what should be matched, but applies it in the speech domain where the query is speech and relevance may depend on both linguistic and paralinguistic cues.

\subsection{Speech/Audio Representation Benchmarks}
Prior work has systematically evaluated speech and audio representations using fixed downstream tasks and probing suites, including NOSS~\cite{shor20_interspeech}, SUPERB~\cite{yang21c_interspeech}, HARES~\cite{wang2022towards}, MSEB~\cite{heigold2025massive}, and MAEB~\cite{assadi2026maeb}.
These benchmarks are highly valuable for measuring whether a representation captures semantic or acoustic information, but they generally do not test instruction-conditioned relevance.
INSPIRE complements them by directly evaluating retrieval behavior under natural-language instructions, where the target attribute can change across queries.
\section{Design of INSPIRE}
\label{sec:design}

\begin{table*}[hbtp]
    \centering
    \caption{Example instructions in INSPIRE. Instructions range from single-attribute criteria to complex multi-attribute combinations.}
    \label{tab:instruction_examples}
    \vspace{-5pt}
    \resizebox{0.9\linewidth}{!}{
        \begin{tabular}{ll}
        \toprule
        \textbf{Instruction Type} & \textbf{Instruction Example}\\
        \midrule
        Continuation & Search for the follow-up lines to the query speech. \\
        Same speaker & Find all documents said by the same talker as the query. \\
        Same speaking style & Return documents with a similar way of speaking to the query. \\
        Same speaker + Same style & Get documents where the voice matches and the speaking style matches. \\
        Same speaker + Sadness style + Same env. & Get sad-style recordings from the same speaker in the same sound environment. \\
        Same speaker + Footsteps env. & Retrieve clips from the same speaker that include footsteps in the background. \\
        \bottomrule
        \end{tabular}
    }
    \vspace{-5pt}
\end{table*}
\begin{table*}[hbtp]
    \centering
    \caption{Each column represents a subtask in subsets, where attribute constraints are defined as follows: S indicates the same value as the query, T indicates a specific target value, and – indicates no constraint.}
    \label{tab:attributes}
    \vspace{-5pt}
    \resizebox{0.9\textwidth}{!}{
    \begin{tabular}{l|c|c|cccc|cccccccccccccc}
        \toprule
        \textbf{Attribute $\backslash$ Subset} 
        & \textbf{DailyTalk} 
        & \textbf{VCTK} 
        & \multicolumn{4}{c|}{\textbf{Expresso}} 
        & \multicolumn{14}{c}{\textbf{Synthetic}} \\
        \midrule
        Semantic & S & -- & -- & -- & -- & -- & S & -- & -- & -- & -- & -- & -- & -- & -- & -- & -- & -- & -- & -- \\
        \midrule
        Speaker & -- & S & S & -- & S & S & -- & S & -- & -- & S & S & S & S & -- & -- & -- & S & S & S \\
        \midrule
        Speaking Style & -- & -- & -- & S & S & T & -- & -- & S & -- & S & T & -- & -- & S & T & S & S & T & S \\
        \midrule
        Environmental Sound & -- & -- & -- & -- & -- & -- & -- & -- & -- & S & -- & -- & S & T & S & S & T & S & S & T \\
        \bottomrule
    \end{tabular}
    }
    \vspace{-10pt}
\end{table*}

\begin{table}[hbtp]
    \centering
    \caption{Statistics of INSPIRE. $|Q|$ denotes the number of unique spoken queries, $|Q_Z|$ the number of query-instruction pairs, $|D|$ the number of documents, $|D^+|/|Q_Z|$ the average number of positive documents per query-instruction pair.}
    \label{tab:statistics}
    \vspace{-5pt}
    \resizebox{0.8\linewidth}{!}{
        \begin{tabular}{lcccc}
            \toprule
            \textbf{Subset} & $|Q|$ & $|Q_Z|$ & $|D|$ & $|D^+|/|Q_Z|$ \\
            \midrule
            DailyTalk  & 200 & 200   & 4{,}882 & 1.000 \\
            Expresso   & 200 & 800   & 3{,}861 & 2.000 \\
            VCTK       & 80  & 80    & 3{,}082 & 2.000 \\
            Synthetic  & 200 & 3{,}000 & 5{,}400 & 1.946 \\
            \midrule
            \textbf{Total / Avg.} & 680 & 4{,}080 & 17{,}225 & 1.910 \\
            \bottomrule
        \end{tabular}
    }
    \vspace{-15pt}
\end{table}

\subsection{Problem Formulation}
\label{subsec:formulation}

We study instruction-aware speech retrieval, in which a system takes a spoken query and a natural-language instruction, then retrieves recordings from a database that satisfy the instruction.

Let $D = \{d_i\}_{i=1}^{|D|}$ denote a speech database with spoken documents and $Q = \{q_i\}_{i=1}^{|Q|}$ a set of spoken queries, where each $d_i$ and $q_i$ is a speech recording whose length may range from a single sentence to a multi-turn passage depending on the retrieval scenario.
Let $Z$ denote the set of all possible natural-language instructions.
For retrieval, we form query-instruction pairs $Q_Z = \{(q, z) \mid q \in Q,\ z \in Z\}$.
In each search, the system receives a spoken query $q$ and an instruction $z$ that specifies the user's intent, potentially constraining the semantic content, speaker identity, vocal style, and acoustic environment.

For any pair $(q, z)$, the database is partitioned into two disjoint sets, $D = D^+ \cup D^-$.
Positive samples $d^+ \in D^+$ satisfy all requirements in $z$ relative to the spoken query $q$, whereas negative samples $d^- \in D^-$ violate at least one requirement.

The instruction-aware speech retrieval system uses an instruction-conditioned scoring function $f(d \mid q, z) \in \mathbb{R}$ to assign a relevance score to each spoken document $d \in D$.
Sorting scores in descending order yields a ranked list $\pi(q, z)$.
Successful retrieval requires $f(d^+ \mid q, z) > f(d^- \mid q, z)$ for all $d^+ \in D^+$ and $d^- \in D^-$, ensuring that relevant spoken documents are ranked above irrelevant ones.

This formulation evaluates how well models interpret natural-language instructions for speech retrieval.
Compared with traditional keyword-matching retrieval, instruction-aware speech retrieval supports free-form instructions that express multi-attribute goals, yielding a more realistic and challenging evaluation setting.

\subsection{Benchmark Construction}
\label{subsec:construction}
INSPIRE consists of four subsets, each built from a distinct data source: DailyTalk~\cite{lee2023dailytalk}, VCTK~\cite{yamagishi2019vctk}, Expresso~\cite{nguyen23_interspeech}, and Synthetic Data. 
Each subset targets a different dimension of instruction-aware retrieval, from conversational continuity to speaker and style matching to complex compositional instructions.
Each subset serves as an independent retrieval corpus, in which queries are evaluated exclusively against the documents within that subset, and all retrieval metrics are computed separately for each subset.
For every subset, we specify (i) the construction of spoken queries, (ii) the relevance criteria, and (iii) the selection of positive and negative documents.

\subsubsection{DailyTalk Subset}
\label{subsubsec:dailytalk}
DailyTalk is a high-quality multi-turn conversational speech dataset designed for modeling contextual dialogue in text-to-speech systems.
This subset evaluates whether a retriever can use a dialogue prefix together with a textual instruction to retrieve the correct continuation.
We use the first half of each dialogue as the spoken query and the second half as the target document.
A document is relevant if it is the true continuation, and negative samples are drawn from unrelated dialogues.

\subsubsection{VCTK Subset}
\label{subsubsec:vctk}
VCTK is a multi-speaker English speech corpus commonly used in speech synthesis and voice conversion studies. 
In this subset, the task is to assess whether a retriever can follow a textual instruction to find utterances spoken by the same speaker as a given spoken query. 
We construct the subset using 80 speakers, where each spoken query serves as a reference utterance, and the goal is to retrieve other utterances from the same speaker. 
Hard negatives are utterances with identical content but spoken by different speakers, while standard negatives differ from the spoken query in both speaker identity and content.

\subsubsection{Expresso Subset}
\label{subsubsec:expresso}
Expresso is a high-quality expressive speech dataset featuring read and improvised dialogues across diverse speaking styles.
This subset evaluates the model's ability to follow a textual instruction to retrieve utterances based on constraints related to speaker, style, or both.
We consider four speakers and five styles: default, whisper, laughing, sad, and confused.
We use the default style only for negative documents.

A document is relevant if it matches the spoken query with respect to the required attributes.
Hard negatives share the same sentence as the spoken query but violate at least one specified attribute, whereas standard negatives differ in both content and attributes.
We exclude same-sentence documents from the positive candidate pool to avoid trivial matches, since these pairs yield high similarity scores that fail to measure the model’s generalization beyond exact matching.
Each query-instruction pair also defines an excluded document set to control the number of positives.

\subsubsection{Synthetic Subset}
\label{subsubsec:synthetic}
We create a synthetic speech retrieval dataset based on Natural Questions~\cite{kwiatkowski2019natural} by converting text queries and documents into speech.
This subset evaluates whether a retrieval model can satisfy multi-attribute constraints specified in text under controlled conditions.
We sample 200 spoken queries and define relevance based on semantic correctness, speaker identity, speaking style, environmental sounds, or their combinations.

Speech is synthesized using the GPT-4o-mini TTS model~\cite{openai2024gpt4ocard} with five distinct voices and three expressive styles: happiness, anger, and sadness.
To simulate realistic acoustic conditions, each utterance is mixed with one environmental sound from the ESC-50~\cite{piczak2015esc} dataset.
We use 15 sound effects such as car horns, rain, thunderstorms, footsteps, keyboard typing, and animal sounds.

Documents are relevant if they satisfy all required attributes.
Hard negatives share the same sentence as the spoken query but violate at least one required attribute, while standard negatives differ in both content and attributes.
Following Expresso, each query-instruction pair defines its own exclusion set to control positive set size and exclude same-content documents from positives, preventing trivial retrieval.

\subsubsection{Instruction Generation}
\label{subsubsec:instruction}
For each relevance type in INSPIRE, we generate 20 diverse instructions using GPT-5.2~\cite{singh2025openaigpt5card}.
Table~\ref{tab:instruction_examples} shows example instructions ranging from simple single-attribute constraints (e.g., ``Find all documents said by the same talker'') to complex multi-attribute combinations (e.g., ``Get sad-style recordings from the same speaker in the same sound environment''). 
For each query-relevance pair, we randomly sample one instruction to form the final retrieval input, promoting linguistic diversity while preserving the underlying retrieval constraints.

\subsection{Benchmark Statistics and Quality Assessments}
\label{subsec:statistics}

Tables~\ref{tab:attributes} and~\ref{tab:statistics} summarize INSPIRE statistics. 
Table~\ref{tab:attributes} shows attribute constraints across subsets. 
DailyTalk focuses exclusively on semantic content matching; VCTK concentrates on speaker identity; Expresso introduces speaking style considerations, with tasks requiring matching speaker alone, style alone, or both simultaneously; and Synthetic incorporates all four attributes: semantic content, speaker identity, speaking style, and environmental sounds. 
For speaking style and environmental acoustics, constraints may require either matching the query or specifying a distinct target value, as these attributes can be controlled independently without a spoken query.
In contrast, semantic content and speaker identity are inherently query-dependent; the former is derived from the transcription of the input speech, while the latter requires a reference utterance to characterize the target voice.

Table~\ref{tab:statistics} reports scale statistics.
The benchmark contains hundreds of spoken queries expanded into thousands of query-instruction pairs via diverse instructions, evaluated against a large document database.
Crucially, the same spoken query paired with different instructions yields different target documents, which is the characteristic of our benchmark.
The low average number of positives per query reflects a realistic scenario in which few relevant items are among many candidates.
DailyTalk provides queries with single positives representing true continuations; VCTK and Expresso offer more constrained evaluations; and Synthetic provides the richest diversity, with multiple instruction types per query reflecting complex multi-attribute matching requirements.

We conduct quality assessments on the synthetic subset across multiple dimensions.
For speech recognition, we apply punctuation filtering and number normalization after transcribing with Whisper-Large~\cite{radford2023robust}.
This yields word error rates of 0.0314 for clean audio and 0.0337 for audio with environmental sounds.
We evaluate speaker consistency using a pre-trained ECAPA-TDNN~\cite{desplanques2020ecapa, ravanelli2021speechbrain} through five-fold SVM~\cite{platt1999probabilistic, chang2011libsvm} cross-validation.
This achieves 0.9998 accuracy in both conditions.
We assess speaking-style quality using emotion embeddings from emotion2vec~\cite{ma2024emotion2vec} with five-fold SVM cross-validation.
This obtains accuracies of 0.8657 for clean audio and 0.8780 for audio with environmental sounds.
We estimate speech quality using UTMOS~\cite{saeki22c_interspeech}, which produces a predicted mean opinion score of 3.76 for clean synthetic speech.
These results demonstrate that the synthetic subset is appropriate for evaluating instruction-aware speech retrieval.
\section{Baseline Methods}
\label{sec:baseline}
We evaluate four baseline approaches that differ in their modality usage and in how they incorporate natural-language instruction.
Each baseline computes a relevance score $f(d \mid q, z)$ given a spoken query $q$, an instruction $z$, and a document $d$.

\subsection{Large Audio-Language Models Embeddings}
\label{subsec:lalm}
Since no existing retriever is designed for instruction-aware speech retrieval, we draw inspiration from work in other modalities that leverage large language models (LLMs)~\cite{jiang2024scaling} or multimodal LLMs (MLLMs)~\cite{jiang2024e5, hu25b_interspeech} as instruction-aware embedding backbones. 
As a baseline, we prompt large audio-language models (LALMs) to encode task-specific instructions into latent representations for retrieval. 
Specifically, for the query side, we concatenate the spoken query~$q$, the instruction~$z$, and the prompt \textit{``Summarize the above sentence and speech in one word.''} as input to the LALM, and extract the query embedding from the hidden state of the last token at layer~$L$: $\mathbf{e}_q = h_{\text{last}}^{(L)}(q, z)$. 
For the document side, we feed the spoken document~$d$ along with the prompt \textit{``Summarize the above speech in one word.''} to obtain $\mathbf{e}_d = h_{\text{last}}^{(L)}(d)$. 
Relevance is then scored via cosine similarity: $f(d \mid q, z) = \cos(\mathbf{e}_q, \mathbf{e}_d)$.

\subsection{Cascaded Pipelines}
\label{subsec:cascaded}
This baseline converts speech inputs into text through automatic speech recognition and captioning, then applies standard text retrieval methods. 
For the spoken query $q$, we generate an ASR transcription $t_q$ that captures the spoken content and a caption $c_q$ that describes acoustic characteristics. 
For each document $d$, we similarly obtain $t_d$ and $c_d$. 
The query and document inputs are constructed as $\tilde{q} = [t_q; c_q; z]$ and $\tilde{d} = [t_d; c_d]$. 
We evaluate both sparse and dense text retrieval approaches. For sparse retrieval, we employ BM25 as $f(d \mid q, z) = \text{BM25}(\tilde{q}, \tilde{d})$. 
For dense retrieval, we encode $\tilde{q}$ and $\tilde{d}$ using a text encoder $\phi(\cdot)$ and compute the relevance score as $f(d \mid q, z) = \cos(\phi(\tilde{q}),\, \phi(\tilde{d}))$.
However, this method relies on the quality of speech-to-text conversion and may lose acoustic information present in the original speech. 
Furthermore, since BM25 and some dense retrievers are not designed to interpret instructions, appending $z$ to the query offers only a superficial form of instruction-awareness rather than a principled solution for instruction-aware retrieval.

\subsection{Self-Supervised Speech Embeddings}
\label{subsec:ssl}
This baseline operates exclusively on speech representations, encoding all utterances without conditioning on the instruction $z$.
We encode both spoken query and document using a self-supervised speech model $g^{(L)}(\cdot)$, where $L$ denotes the layer from which representations are extracted, as $\mathbf{e}_q = g^{(L)}(q)$ and $\mathbf{e}_d = g^{(L)}(d)$. 
The relevance score is computed as $f(d \mid q, z) = \cos(\mathbf{e}_q, \mathbf{e}_d)$, which is independent of the instruction $z$.
Since this approach has no mechanism to condition retrieval on the given instruction, it is inherently unable to distinguish between different instructions for the same spoken query. 
As such, it is not expected to perform well on instruction-aware retrieval tasks. 
We include it primarily as a baseline to quantify the gains achieved by instruction-conditioned methods.

\subsection{Contrastive Audio-Language Embeddings}
\label{subsec:clap}
This baseline uses a contrastive audio-language model, such as CLAP~\cite{elizalde2023clap, wu2023large}, for cross-modal text-audio retrieval. 
The spoken query is converted into text and concatenated with the instruction as $\tilde{q} = [t_q; c_q; z]$, then encoded using the text encoder $\psi_{\text{text}}(\cdot)$, while each document is encoded directly from speech using the audio encoder $\psi_{\text{audio}}(\cdot)$. 
The relevance score is computed as $f(d \mid q, z) = \cos\big(\psi_{\text{text}}(\tilde{q}),\, \psi_{\text{audio}}(d)\big)$. 
Although this preserves acoustic information in document representations, the encoders are not trained to ground $z$ in the joint embedding space, so concatenating $z$ is unlikely to modulate retrieval meaningfully. 
We include this baseline to assess the models' zero-shot performance on our task.

\subsection{Reranking}
\label{subsec:rerank_method}
We also experiment with several reranking methods. 
Given the top-$K$ documents retrieved by the first-stage retriever, the reranker reassigns a relevance score to each candidate and reorders them accordingly. 
Drawing inspiration from work in other modalities that leverage LLMs or MLLMs as rerankers~\cite{zhang2025qwen3, li2026qwen3}, we explore LALM-based rerankers that take as input the instruction $z$, spoken query $q$, and candidate document $d$, along with the prompt \textit{``Judge whether the Document meets the requirements based on the Query and the Instruct provided. Note that the answer can only be `yes' or `no'.''}. 
The reranking score is computed as $\sigma(\mathrm{logit}(\text{yes}) - \mathrm{logit}(\text{no}))$. 
In addition, we explore text-based rerankers, where the spoken query and document are replaced by their respective transcriptions and captions $[t_q; c_q]$ and $[t_d; c_d]$, while retaining the same instruction $z$, prompt, and scoring function.
\setlength{\abovedisplayskip}{3pt}
\setlength{\belowdisplayskip}{3pt}
\section{Experiments}
\label{sec:experiments}
\begin{table*}[hbtp]
    \centering
    \caption{Main results on INSPIRE. We report Recall@10/50/100 across four subsets. Model size is reported in billions of parameters, and the best results per subset are shown in bold.}
    \label{tab:main}
    \vspace{-5pt}
    \resizebox{0.98\textwidth}{!}{
    \begin{tabular}{ll|c|ccc|ccc|ccc|ccc}
        \toprule
        & & \multirow{2}{*}{\textbf{Size (B)}} & \multicolumn{3}{c|}{\textbf{DailyTalk}} & \multicolumn{3}{c|}{\textbf{VCTK}} & \multicolumn{3}{c|}{\textbf{Expresso}} & \multicolumn{3}{c}{\textbf{Synthetic}} \\
        & & & \textbf{R@10} & \textbf{R@50} & \textbf{R@100} & \textbf{R@10} & \textbf{R@50} & \textbf{R@100} & \textbf{R@10} & \textbf{R@50} & \textbf{R@100} & \textbf{R@10} & \textbf{R@50} & \textbf{R@100} \\
        \midrule
        \multicolumn{15}{c}{\textit{Random Baseline}} \\
        \midrule
        Random & & - & 0.24 & 1.06 & 2.10 & 0.35 & 1.84 & 3.63 & 0.30 & 1.50 & 2.96 & 0.20 & 1.00 & 1.98 \\
        \midrule
        \multicolumn{15}{c}{\textit{LALMs Embeddings}} \\
        \midrule
        Audio-Flamingo-3~\cite{goel2025audio} & & 8 & 30.00 & 57.50 & 69.50 & 1.25 & 6.25 & 7.50 & 2.19 & 7.31 & 10.94 & 3.82 & 6.31 & 7.72 \\
        Qwen2.5-Omni~\cite{xu2025qwen2} & & 3 & 26.50 & 53.00 & 64.00 & 1.88 & 3.13 & 3.13 & 0.75 & 3.75 & 6.69 & 3.14 & 5.20 & 6.88 \\
        Qwen2.5-Omni~\cite{xu2025qwen2} & & 7 & 36.00 & 62.50 & 73.00 & 1.25 & 3.13 & 5.63 & 0.63 & 3.38 & 6.13 & 4.18 & 6.00 & 7.47 \\
        Qwen3-Omni~\cite{xu2025qwen3} & & 30 & 51.50 & 74.50 & 83.00 & 1.25 & 3.13 & 6.25 & 0.94 & 3.56 & 6.19 & 4.88 & 6.38 & 7.88 \\
        Voxtral-Mini~\cite{liu2025voxtral} & & 3 & 39.00 & 59.00 & 68.00 & 0.63 & 2.50 & 5.63 & 0.38 & 1.38 & 3.13 & 3.87 & 5.40 & 7.20 \\
        Voxtral-Small~\cite{liu2025voxtral} & & 24 & 41.00 & 67.50 & 78.00 & 1.25 & 3.75 & 5.63 & 0.31 & 1.44 & 2.81 & 5.17 & 7.23 & 8.38 \\
        \midrule
        \multicolumn{15}{c}{\textit{Cascaded Pipelines}} \\
        \midrule
        BM25~\cite{robertson2009probabilistic} & & - & 36.50 & 59.50 & 65.50 & 1.25 & 5.63 & 6.88 & 0.56 & 3.06 & 6.19 & 4.06 & 6.89 & 8.95 \\
        SentenceBERT~\cite{reimers2019sentence} & & 0.1 & 32.50 & 50.50 & 59.50 & 0.63 & 3.75 & 4.38 & 0.44 & 2.63 & 5.13 & 4.16 & 6.28 & 7.92 \\
        E5-Mistral~\cite{wang2024improving} & & 7 & 55.50 & 78.00 & 85.00 & 1.25 & 5.63 & 10.63 & 1.63 & 5.25 & 8.31 & 6.50 & 8.32 & 9.92 \\
        Qwen3-Embedding~\cite{zhang2025qwen3} & & 8 & \textbf{62.00} & \textbf{81.50} & \textbf{89.50} & 1.88 & 6.25 & 10.63 & 0.75 & 5.56 & 9.69 & \textbf{7.02} & \textbf{8.95} & \textbf{11.07} \\
        \midrule
        \multicolumn{15}{c}{\textit{Self-Supervised Speech Embeddings}} \\
        \midrule
        HuBERT-Large~\cite{hsu2021hubert} & & 0.3 & 11.00 & 18.00 & 26.50 & 11.88 & 31.88 & 37.50 & \textbf{9.81} & \textbf{19.44} & \textbf{24.19} & 1.28 & 3.70 & 5.70 \\
        WavLM-Large~\cite{chen2022wavlm} & & 0.3 & 16.50 & 24.50 & 35.50 & \textbf{17.50} & \textbf{35.00} & \textbf{43.75} & 9.50 & 17.69 & 24.00 & 1.70 & 4.48 & 7.22 \\
        \midrule
        \multicolumn{15}{c}{\textit{Contrastive Audio-Language Embeddings}} \\
        \midrule
        LAION-CLAP~\cite{wu2023large} & & 0.2 & 0.00 & 1.50 & 2.00 & 1.94 & 5.63 & 8.94 & 1.88 & 4.38 & 7.50 & 0.80 & 2.22 & 4.35 \\
        \bottomrule
    \end{tabular}
    }
    \vspace{-10pt}
\end{table*}

\subsection{Experiment Setup}
\label{subsec:setup}
Following Section~\ref{sec:baseline}, we evaluate four baseline retrieval paradigms and several reranking methods.

For LALM embedding, we evaluate six LALMs spanning different parameter sizes and architectures: Audio-Flamingo-3~\cite{goel2025audio}, Qwen2.5-Omni-3B/7B~\cite{xu2025qwen2}, Qwen3-Omni-30B-A3B-Instruct~\cite{xu2025qwen3}, Voxtral-Mini-3B, and Voxtral-Small-24B~\cite{liu2025voxtral}. 
Following prior work~\cite{jiang2024scaling, jiang2024e5, hu25b_interspeech}, we extract embeddings from the hidden states of the last layer.
For the cascaded pipeline, we use Whisper-large-v3~\cite{radford2023robust} for ASR and Qwen3-Omni-30B-A3B-Captioner~\cite{xu2025qwen3} for captioning, with BM25~\cite{robertson2009probabilistic} as a sparse retriever, SentenceBERT~\cite{reimers2019sentence} as a dense retriever, and E5-Mistral~\cite{wang2024improving} and Qwen3-Embedding-8B~\cite{zhang2025qwen3} as instruction-aware dense retrievers. 
For self-supervised speech embedding, we evaluate HuBERT-Large~\cite{hsu2021hubert} and WavLM-Large~\cite{chen2022wavlm} using last-layer representations. 
For contrastive audio-language embeddings, we employ LAION-CLAP~\cite{wu2023large}. 
For reranking, we use Qwen2.5-Omni-3B and Voxtral-Mini-3B as LALM rerankers, and Qwen3-Reranker-0.6B and 4B as text rerankers.  
We opt for slightly smaller models because the reranker scores each query-document pair individually, making it far more computationally expensive than the retriever.

We use recall as the primary retrieval metric. 
For a query-instruction pair $(q, z)$, let $\pi_K(q, z)$ denote the top-$K$ ranked documents and $\mathcal{D}^+$ the set of positive documents. Recall@$K$ is defined as: 
\begin{equation} 
\mathrm{Recall@}K = \frac{1}{|Q_Z|} \sum_{(q, z)\in Q_Z} \frac{\left| \pi_K(q, z) \cap \mathcal{D}^+ \right|}{\left| \mathcal{D}^+ \right|}, \end{equation} 
where $Q_Z$ denotes the set of query-instruction pairs. 
For reranking, we adopt NDCG, which assigns greater credit to relevant documents at higher positions: \begin{equation} 
\text{NDCG@K} = \frac{\text{DCG@K}}{\text{IDCG@K}}, \hspace{0.5em} \text{DCG@K} = \sum_{i=1}^{K} \frac{2^{r_i} - 1}{\log_2(i+1)} 
\end{equation} 
where $r_i$ is the relevance label of the document at rank $i$, and IDCG@$K$ is the ideal DCG from sorting by relevance.

\subsection{Main Results}
\label{subsec:main}
Table~\ref{tab:main} shows retrieval performance across all INSPIRE subsets. 
We also include a random baseline that uniformly samples documents, averaged over 100 runs, confirming that higher results are meaningful.

On DailyTalk, which targets semantic retrieval of conversational continuity, cascaded pipelines achieve the best performance.
Instruction-aware dense retrievers such as Qwen3-Embedding obtain the strongest results, followed closely by E5-Mistral. 
LALMs also perform competitively but fall short of the performance of cascaded approaches. 
Contrastive audio-language models underperform due to weak semantic alignment between text and speech. 
These results suggest that converting speech to text and leveraging powerful text embedding models remain the most effective strategies for semantic retrieval.

On VCTK and Expresso, which emphasize speaker identity and speaking style, self-supervised speech models exhibit clear advantages.
These subsets involve simpler instructions in which retrieval depends on paralinguistic features, expressiveness, prosody, emotion, and speaker traits, which are better captured acoustically. 
HuBERT-Large and WavLM-Large outperform most other approaches, showing that direct speech representations preserve more relevant information than cascaded methods for acoustic and paralinguistic properties.

On the Synthetic subset, which combines multiple attributes spanning semantic content, speaker characteristics, speaking style, and environmental sounds, the task proves especially challenging.
Self-supervised speech embeddings achieve very low performance, confirming that instruction-agnostic representations struggle with multi-attribute retrieval.
Contrastive audio-language embeddings also underperform due to diverse scenarios. 
LALMs and cascaded pipelines perform better but still yield relatively low results, suggesting this subset is a genuinely challenging benchmark requiring joint understanding of complex instructions and diverse acoustic properties.

\begin{figure*}[t]
    \centering
    \begin{subfigure}{0.33\linewidth}
        \centering
        \includegraphics[width=\linewidth]{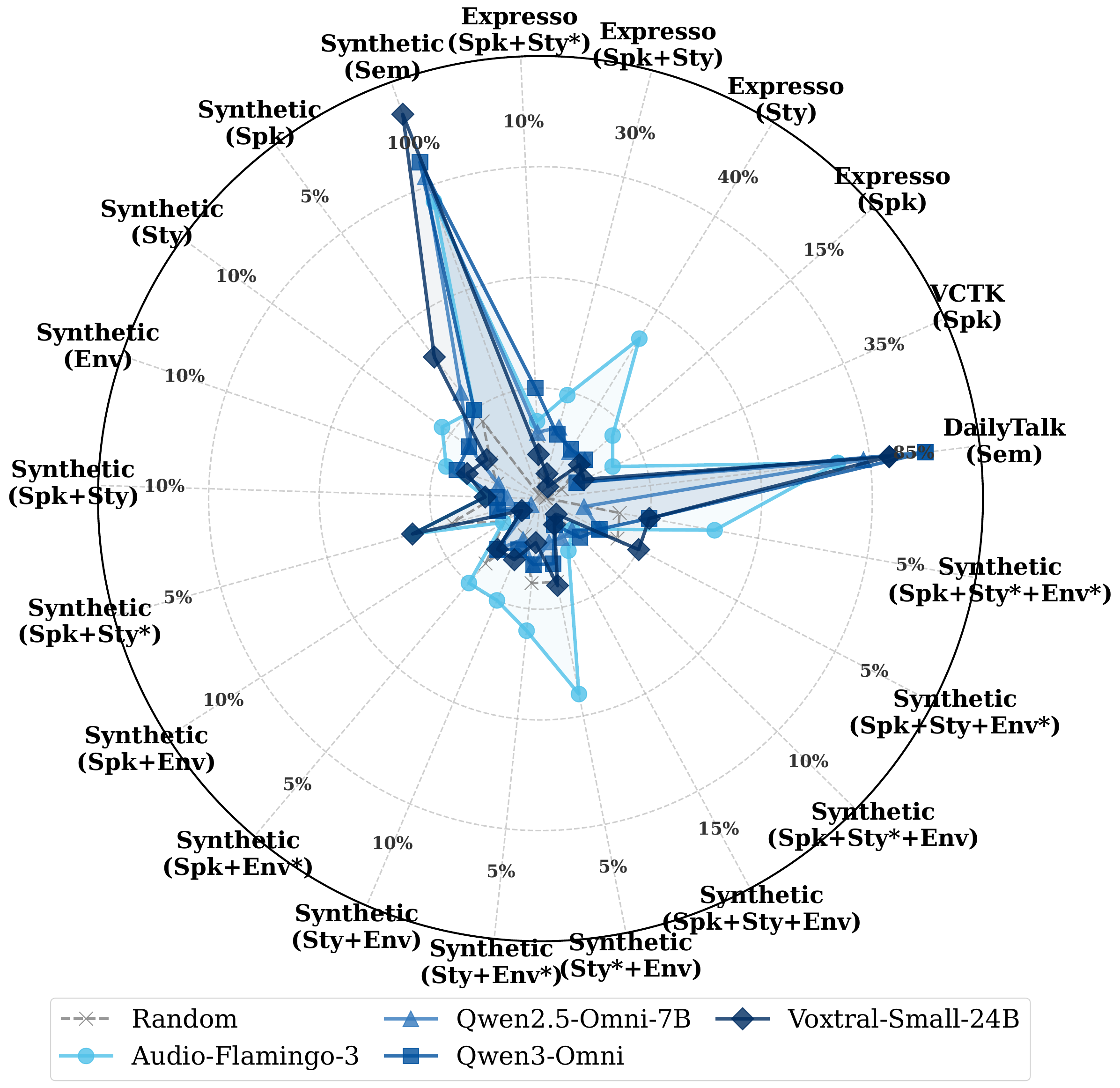}
    \end{subfigure}
    \begin{subfigure}{0.33\linewidth}
        \centering
        \includegraphics[width=\linewidth]{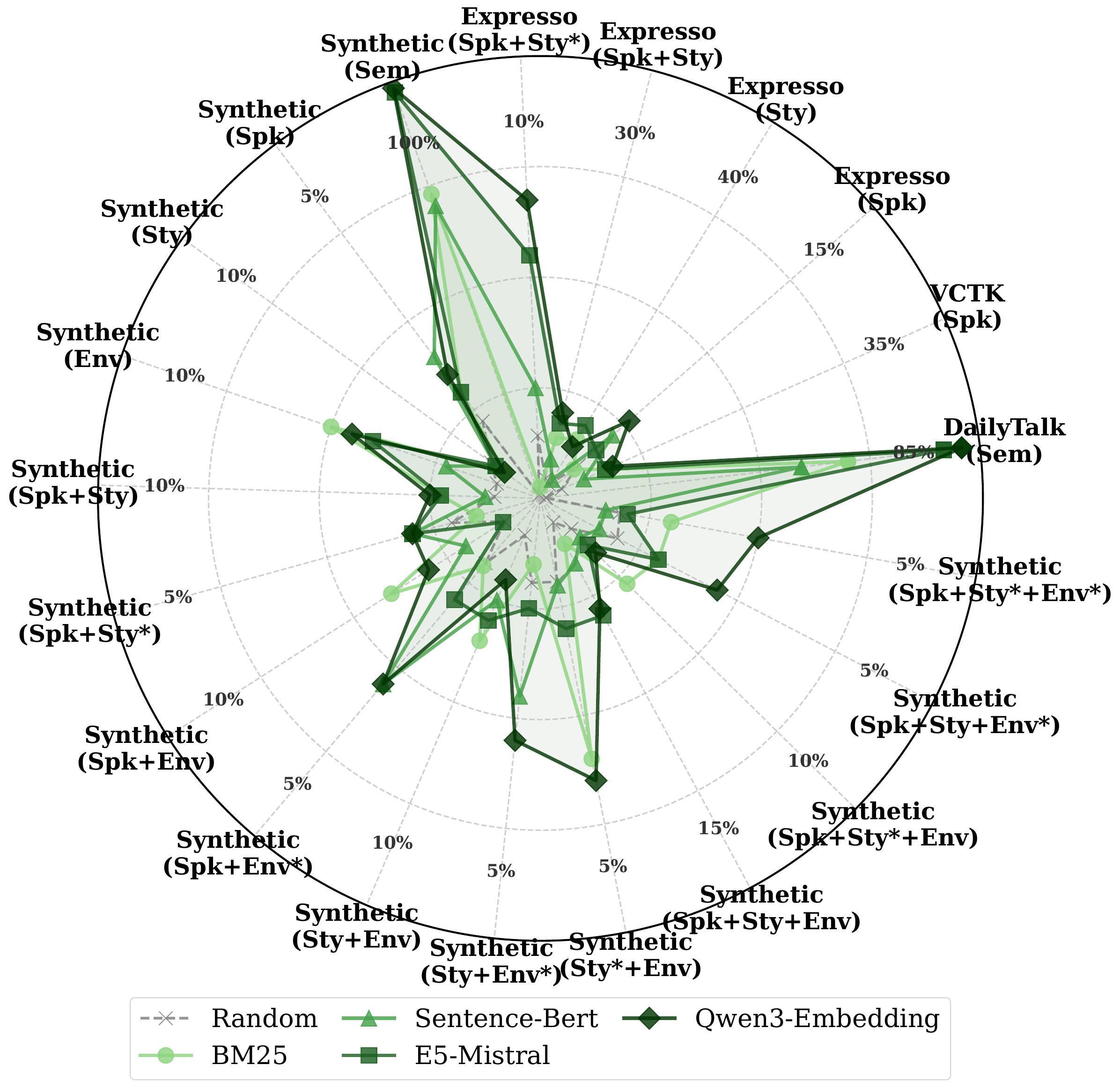}
    \end{subfigure}
    \begin{subfigure}{0.33\linewidth}
    \centering
    \includegraphics[width=\linewidth]{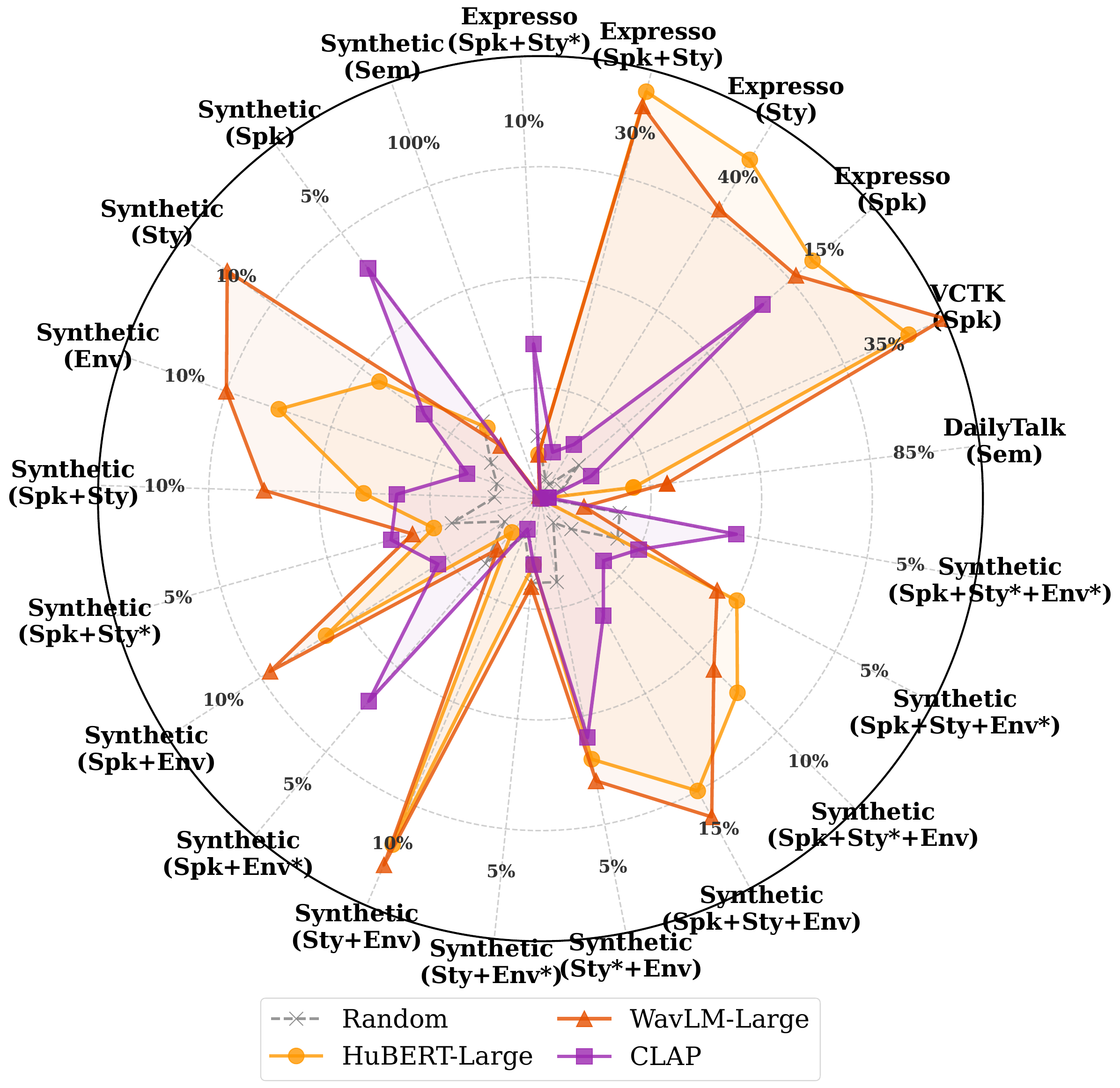}
    \end{subfigure}
    \caption{Performance across attribute constraints shown in radar plots for LALMs (left), cascaded pipelines (middle), and self-supervised/contrastive models (right). 
    We report Recall@50. 
    Sem refers to semantic, Spk to speaker, Sty to speaking style, and Env to environmental sound.
    Attributes marked with * specify a distinct target value rather than matching the query.
    Scores are normalized by the best-performing model per attribute, with a shared scale across all subfigures.}
    \label{fig:instruction_types}
    \vspace{-10pt}
\end{figure*}

In summary, our results reveal distinct strengths and limitations across different retrieval paradigms. 
All methods suffer significant performance degradation in complex multi-attribute scenarios, indicating substantial room for improvement in handling compositional instructions. 
Different approaches excel along different retrieval dimensions depending on whether the task prioritizes semantic or acoustic features. 
More broadly, our benchmark reveals a fundamental trade-off between semantic understanding, where cascaded methods prevail, and acoustic feature preservation, where speech-based methods excel. 
These findings highlight the need for unified models that effectively integrate both modalities, enabling robust performance across diverse retrieval intents while bridging the gap between semantic comprehension and acoustic representation.

\subsection{Reranking Results}
\label{subsec:reranking}
\begin{table}[hbtp]
    \centering
    \caption{Reranking performance measured by NDCG@10/50 on the top 100 documents from first-stage retrievers.}
    \vspace{-5pt}
    \label{tab:rerank}
    \resizebox{\linewidth}{!}{
    \begin{tabular}{lcccccccc}
        \toprule
        & \multicolumn{2}{c}{\textbf{DailyTalk}} & \multicolumn{2}{c}{\textbf{VCTK}} & \multicolumn{2}{c}{\textbf{Expresso}} & \multicolumn{2}{c}{\textbf{Synthetic}} \\
        \cmidrule(lr){2-3} \cmidrule(lr){4-5} \cmidrule(lr){6-7} \cmidrule(lr){8-9}
        \textbf{Model} & \textbf{N@10} & \textbf{N@50} & \textbf{N@10} & \textbf{N@50} & \textbf{N@10} & \textbf{N@50} & \textbf{N@10} & \textbf{N@50} \\
        \midrule
        \multicolumn{9}{c}{\textit{LALMs Embeddings Rerank with LALMs}} \\
        \midrule
        Qwen2.5-Omni-7B & 21.47 & 27.45 & 1.10 & 1.52 & 0.46 & 1.18 & 2.02 & 2.47 \\
        + Qwen2.5-Omni-3B & 38.50 & 41.69 & 0.77 & 1.34 & 1.05 & 1.53 & 3.29 & 3.47 \\
        + Voxtral-Mini-3B & 46.32 & 48.32 & 1.01 & 1.79 & 0.18 & 0.77 & 3.14 & 3.43 \\
        Voxtral-Small & 26.60 & 32.55 & 1.53 & 2.17 & 0.12 & 0.41 & 2.59 & 3.08 \\
        + Qwen2.5-Omni-3B & 37.17 & 40.80 & 1.04 & 1.62 & 0.58 & 0.88 & 3.52 & 3.79 \\
        + Voxtral-Mini-3B & 45.72 & 48.33 & 1.01 & 1.66 & 0.11 & 0.40 & 3.36 & 3.70 \\
        \midrule
        \multicolumn{9}{c}{\textit{Cascaded Pipelines Rerank with Text Rerankers}} \\
        \midrule
        BM25 & 25.35 & 30.68 & 1.25 & 2.36 & 0.21 & 0.87 & 2.02 & 2.73 \\
        + Qwen3-Reranker-0.6B & 22.09 & 26.90 & 1.32 & 2.29 & 0.31 & 1.15 & 1.57 & 2.38 \\
        + Qwen3-Reranker-4B & 24.77 & 29.05 & 0.97 & 1.29 & 0.37 & 0.94 & 1.27 & 2.11 \\
        Qwen3-Embedding & 43.05 & 47.39 & 1.48 & 2.54 & 0.34 & 1.55 & 3.58 & 4.07 \\
        + Qwen3-Reranker-0.6B & 23.27 & 31.20 & 1.63 & 2.96 & 0.39 & 1.53 & 2.02 & 3.09 \\
        + Qwen3-Reranker-4B & 28.47 & 35.43 & 0.90 & 2.11 & 0.43 & 1.45 & 2.03 & 3.07 \\
        \bottomrule
    \end{tabular}
    }
    \vspace{-15pt}
\end{table}
We evaluate reranking performance on top of the first-stage retrievers. For each query-instruction pair, the top-100 retrieved documents are reranked.
For the LALM retrievers, we apply LALM-based rerankers. 
For the cascaded pipeline, we employ text rerankers instead.

As shown in Table~\ref{tab:rerank}, LALM reranking generally outperforms the first-stage LALM retrievers, with particularly great improvements on DailyTalk. 
In contrast, text-based reranking substantially improves performance on DailyTalk, while achieving comparable results on the other subsets.
We hypothesize that this phenomenon arises because concatenating transcriptions and captions introduces excessive information that may mislead the reranker. 
Additionally, the instruction formats in INSPIRE are likely out of domain for text rerankers trained on standard retrieval tasks.
These findings suggest that direct access to speech signals benefits LALM rerankers, whereas text rerankers require in-domain instruction-aware fine-tuning to perform effectively in this setting.

\section{Ablations and Analysis}
\label{sec:analysis}

\subsection{Analysis of Instruction Types}
\label{subsec:instruction_types}

To better understand which parts of an instruction current retrievers can handle, we break down performance by instruction type, corresponding to different attribute constraints.
Figure~\ref{fig:instruction_types} reports a fine-grained comparison among LALMs, cascaded pipelines, self-supervised speech models, and contrastive audio-language models.
To ensure visual clarity across tasks of varying difficulty, each radar axis is normalized by the best-performing model on that attribute.

The results reveal a clear modality specialization.
Cascaded methods consistently perform best on semantically driven instructions.
For semantic continuation and answer containment, they substantially outperform random baselines and LALMs, suggesting that specialized text embeddings remain more robust for precise semantic correspondence in retrieval.

In contrast, the trend reverses for acoustic attribute matching.
Self-supervised speech models excel at satisfying same-attribute constraints related to speaker identity, speaking style, and acoustic environment, whereas the cascaded method and LALMs frequently perform at near-chance levels on these dimensions.
This performance gap indicates that current text-centric and unified multimodal architectures lack sufficient ability to disentangle paralinguistic cues from linguistic content.

Finally, when instructions specify particular attribute values, self-supervised speech models show notably weaker performance, suggesting that fine-grained control in instruction-aware speech retrieval still requires more targeted supervision or attribute-aware training strategies.

\subsection{Impact of Instruction Usage}
\label{subsec:instruction_usage}
\begin{figure}[hbtp]
    \centering
    \vspace{-10pt}
    \includegraphics[width=\linewidth]{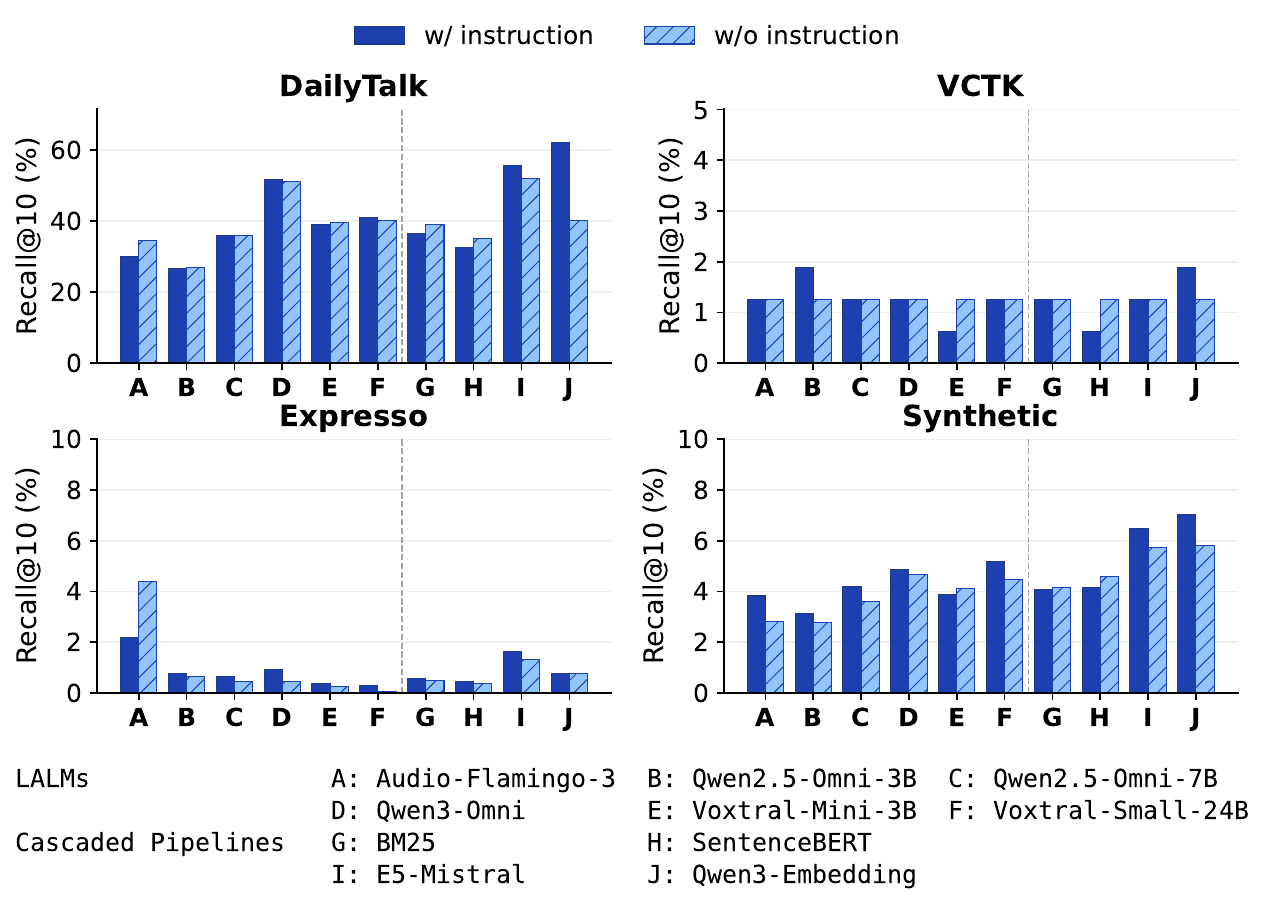}
    \vspace{-15pt}
    \caption{Impact of instruction usage on retrieval performance. 
    Solid bars show results with instructions; hatched bars without. 
    Models A to F are LALMs; G to J are cascaded pipelines.}
    \label{fig:instruction_usage}
    \vspace{-5pt}
\end{figure}
To assess whether models effectively leverage instructions, we compare retrieval performance with and without instructions. 
Figure~\ref{fig:instruction_usage} shows results for various models under both conditions across all four subsets.

Overall, instruction gains are evident only in instruction-aware text retrievers such as E5-Mistral and Qwen3-Embedding, while BM25 and SentenceBERT show minimal sensitivity due to their instruction-agnostic design. 
All six LALMs also exhibit limited instruction sensitivity, as they lack instruction-aware retrieval training.

On DailyTalk, instruction-aware text retrievers show substantial gains when instructions are provided. 
On VCTK, Expresso, and Synthetic, all models show minimal differences between conditions, indicating current approaches cannot leverage instructions for paralinguistic and multi-attribute retrieval.

These results suggest that instruction-aware training is necessary for models to benefit from instructions.

\subsection{Comparison of Captioning Models}
\label{subsec:captioning_models}
\begin{figure}[hbtp]
    \centering    
    \vspace{-10pt}
    \includegraphics[width=\linewidth]{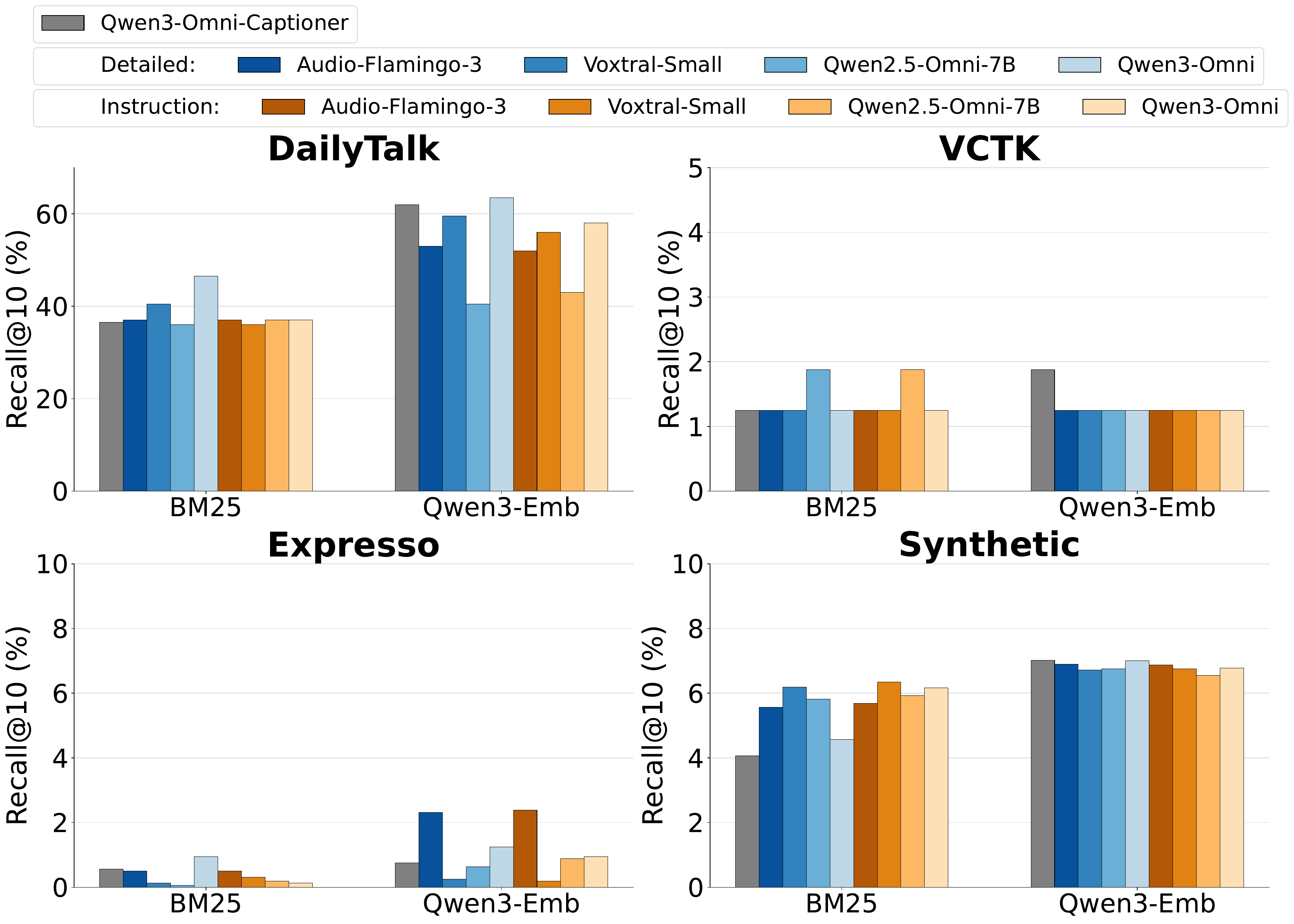}
    \vspace{-15pt}
    \caption{Comparison of captioning models for cascaded pipelines using detailed and instruction-aware prompts.}
    \label{fig:captioner}
    \vspace{-5pt}
\end{figure}
Figure~\ref{fig:captioner} reports retrieval performance in a controlled setting where only the captioner is varied while the downstream text retriever is held fixed, isolating the effect of caption quality from retriever choice.
We also conduct prompt engineering ablations for the other LALMs, since the Qwen3-Omni-Captioner does not accept text input. 
Specifically, we prompt the other LALMs with a detailed prompt, \textit{``Generate a caption that describes the speech above, including its meaning, the speaker's identity or role, the speaking style, and any relevant environmental sounds.''}, and an instruction-aware prompt, \textit{``Generate a caption that describes the speech above and can be used for the following instruction: \{instruction\}''}.

The detailed and instruction-aware prompts yield nearly identical results across most configurations. 
Furthermore, all evaluated methods exhibit a sharp performance floor on the VCTK and Expresso datasets. 
The consistently low scores on these benchmarks, regardless of the LALM or prompting strategy employed, suggest that these tasks remain inherently challenging and are currently less sensitive in the caption model, likely due to a misalignment between the speech features and the retrieval objectives.

\subsection{Effect of Oracle Metadata}
\label{subsec:oracle}
\begin{table}[hbtp]
    \centering
    \definecolor{darkgreen}{rgb}{0.0, 0.6, 0.0}
    \caption{Effect of oracle metadata on cascaded text retrieval. \textcolor{darkgreen}{Green} indicates improvement and \textcolor{red}{red} indicates degradation compared to the caption baseline.}
    \label{tab:oracle}
    \vspace{-5pt}
    \resizebox{\linewidth}{!}{
        \begin{tabular}{lcccccccc}
        \toprule
        & \multicolumn{2}{c}{\textbf{DailyTalk}} & \multicolumn{2}{c}{\textbf{VCTK}} & \multicolumn{2}{c}{\textbf{Expresso}} & \multicolumn{2}{c}{\textbf{Synthetic}} \\
        \cmidrule(lr){2-3} \cmidrule(lr){4-5} \cmidrule(lr){6-7} \cmidrule(lr){8-9}
        \textbf{Model} & \textbf{R@10} & \textbf{R@50} & \textbf{R@10} & \textbf{R@50} & \textbf{R@10} & \textbf{R@50} & \textbf{R@10} & \textbf{R@50} \\
        \midrule
        BM25 & 36.50 & 59.50 & 1.25 & 5.63 & 0.56 & 3.06 & 4.06 & 6.89 \\
        + Oracle & \textcolor{red}{36.00} & \textcolor{red}{49.00} & \textcolor{darkgreen}{53.13} & \textcolor{darkgreen}{95.00} & \textcolor{darkgreen}{0.63} & \textcolor{red}{2.69} & \textcolor{darkgreen}{5.18} & \textcolor{darkgreen}{7.57} \\
        \midrule
        SentenceBERT & 32.50 & 50.50 & 0.63 & 3.75 & 0.44 & 2.63 & 4.16 & 6.28 \\
        + Oracle & \textcolor{red}{14.00} & \textcolor{red}{25.50} & 0.63 & \textcolor{red}{1.25} & \textcolor{darkgreen}{0.50} & \textcolor{darkgreen}{3.50} & \textcolor{red}{1.53} & \textcolor{red}{3.86} \\
        \midrule
        E5-Mistral & 55.50 & 78.00 & 1.25 & 5.63 & 1.63 & 5.25 & 6.50 & 8.32 \\
        + Oracle & \textcolor{red}{49.00} & \textcolor{red}{71.00} & \textcolor{darkgreen}{1.88} & \textcolor{red}{4.38} & \textcolor{red}{1.19} & \textcolor{darkgreen}{5.75} & \textcolor{darkgreen}{7.12} & \textcolor{darkgreen}{10.23} \\
        \midrule
        Qwen3-Embedding & 62.00 & 81.50 & 1.88 & 6.25 & 0.75 & 5.56 & 7.02 & 8.95 \\
        + Oracle & \textcolor{red}{58.00} & \textcolor{red}{80.50} & 1.88 & \textcolor{darkgreen}{8.13} & \textcolor{darkgreen}{2.69} & \textcolor{darkgreen}{9.50} & \textcolor{darkgreen}{7.52} & \textcolor{darkgreen}{10.20} \\
        \bottomrule
        \end{tabular}
    }
    \vspace{-10pt}
\end{table}
To investigate the impact of metadata quality on cascaded pipelines, we replace model-generated captions with oracle metadata containing ground-truth transcriptions, speaker IDs, speaking styles, and environmental sound labels.
Table~\ref{tab:oracle} summarizes the results.

Oracle metadata degrades semantic retrieval on DailyTalk, as non-semantic attributes introduce noise that misleads text matching. 
On paralinguistic subsets such as VCTK and Expresso, effects are mixed but often positive. 
On VCTK, BM25 benefits dramatically because oracle speaker IDs provide highly distinctive tokens ideal for term matching. 
Oracle labels also help instruction-aware embeddings when retrieval targets these attributes directly.
On Synthetic, oracle metadata improves most models but not uniformly, highlighting that even perfect metadata cannot ensure robust multi-attribute retrieval. 
These findings indicate that both accurate metadata extraction and instruction-aware training are necessary for effective cascaded retrieval on non-semantic tasks.

\subsection{Comparison with Proprietary Models}
\label{subsec:proprietary}
\begin{table}[hbtp]
    \centering
    \vspace{-10pt}
    \caption{Comparison with proprietary models.
    Results are shown for representative open-source models and proprietary APIs from Google and OpenAI.}
    \label{tab:proprietary}
    \vspace{-5pt}
    \resizebox{\linewidth}{!}{
    \begin{tabular}{lcccccccc}
        \toprule
        & \multicolumn{2}{c}{\textbf{DailyTalk}} & \multicolumn{2}{c}{\textbf{VCTK}} & \multicolumn{2}{c}{\textbf{Expresso}} & \multicolumn{2}{c}{\textbf{Synthetic}} \\
        \cmidrule(lr){2-3} \cmidrule(lr){4-5} \cmidrule(lr){6-7} \cmidrule(lr){8-9}
        \textbf{Model} & \textbf{R@10} & \textbf{R@50} & \textbf{R@10} & \textbf{R@50} & \textbf{R@10} & \textbf{R@50} & \textbf{R@10} & \textbf{R@50} \\
        \midrule
        \multicolumn{9}{c}{\textit{LALMs Embeddings}} \\
        \midrule
        Qwen3-Omni & 51.50 & 74.50 & 1.25 & 3.13 & 0.94 & 3.56 & 4.88 & 6.38 \\
        Voxtral-Small & 41.00 & 67.50 & 1.25 & 3.75 & 0.31 & 1.44 & 5.17 & 7.23 \\
        \midrule
        \multicolumn{9}{c}{\textit{Cascaded Pipelines with Open-source Models}} \\
        \midrule
        E5-Mistral & 55.50 & 78.00 & 1.25 & 5.63 & 1.63 & 5.25 & 6.50 & 8.32 \\
        Qwen3-Embedding & 62.00 & 81.50 & 1.88 & 6.25 & 0.75 & 5.56 & 7.02 & 8.95 \\
        \midrule
        \multicolumn{9}{c}{\textit{Cascaded Pipelines with Proprietary Models}} \\
        \midrule
        Gemini & 53.00 & 80.00 & 1.25 & 3.13 & 0.81 & 3.13 & 6.97 & 8.40 \\
        OpenAI & 57.50 & 80.00 & 1.25 & 1.25 & 0.06 & 1.38 & 6.73 & 7.50 \\
        \bottomrule
        \end{tabular}
    }
    \vspace{-5pt}
\end{table}
We compare proprietary APIs from Google and OpenAI in Table~\ref{tab:proprietary}. 
Since no proprietary multimodal speech embedding model currently exists, we follow the same cascaded pipeline (Section~\ref{subsec:cascaded}): captioning with a proprietary MLLM and embedding with a proprietary text model.
For Google, we use Gemini-3.0-Flash~\cite{gemini-3-pro} and gemini-embedding-001~\cite{lee2025gemini}; for OpenAI, GPT-4o-mini-Audio~\cite{openai2024gpt4ocard} and text-embedding-3-large~\cite{openaiemb}. 

Proprietary models are competitive on semantic retrieval but do not consistently outperform the best open-source instruction-aware text retrievers. 
On DailyTalk, both trail the strongest instruction-aware embedding model, suggesting that caption quality alone does not fully close the semantic retrieval gap.
On VCTK and Expresso, both underperform the best open-source baselines, further confirming the limitations of cascaded pipelines under paralinguistic constraints.
On Synthetic, results are comparable to the best open-source baselines, though overall recall remains low.

These findings highlight a fundamental limitation of caption-then-embed pipelines, as compressing rich speech into text often drops fine-grained acoustic cues, and further motivate instruction-aware speech representations or multimodal embeddings that directly encode acoustic attributes.

\subsection{LALMs Pooling Strategy Comparison}
\label{subsec:lalm_pooling}
\begin{table}[hbtp]
    \centering
    \vspace{-10pt}
    \caption{LALMs pooling strategy comparison. We compare the prompt-based last token embedding approach versus mean pooling without prompts.}
    \label{tab:pooling}
    \vspace{-5pt}
    \resizebox{\linewidth}{!}{
    \begin{tabular}{lcccccccc}
        \toprule
        & \multicolumn{2}{c}{\textbf{DailyTalk}} & \multicolumn{2}{c}{\textbf{VCTK}} & \multicolumn{2}{c}{\textbf{Expresso}} & \multicolumn{2}{c}{\textbf{Synthetic}} \\
        \cmidrule(lr){2-3} \cmidrule(lr){4-5} \cmidrule(lr){6-7} \cmidrule(lr){8-9}
        \textbf{Model} & \textbf{R@10} & \textbf{R@50} & \textbf{R@10} & \textbf{R@50} & \textbf{R@10} & \textbf{R@50} & \textbf{R@10} & \textbf{R@50} \\
        \midrule
        Audio-Flamingo-3 & 30.00 & 57.50 & 1.25 & 6.25 & 2.19 & 7.31 & 3.82 & 6.31 \\
        + Mean Pooling & 34.00 & 56.50 & 3.75 & 8.13 & 1.94 & 6.25 & 0.57 & 2.08 \\
        \midrule
        Qwen2.5-Omni-7B & 36.00 & 62.50 & 1.25 & 3.13 & 0.63 & 3.38 & 4.18 & 6.00 \\
        + Mean Pooling & 27.00 & 44.50 & 0.63 & 2.50 & 1.50 & 4.25 & 1.22 & 3.68 \\
        \midrule
        Qwen3-Omni & 51.50 & 74.50 & 1.25 & 3.13 & 0.94 & 3.56 & 4.88 & 6.38 \\
        + Mean Pooling & 15.00 & 25.50 & 1.25 & 5.63 & 0.81 & 3.56 & 0.45 & 1.63 \\
        \midrule
        Voxtral-Small-24B & 41.00 & 67.50 & 1.25 & 3.75 & 0.31 & 1.44 & 5.17 & 7.23 \\
        + Mean Pooling & 28.00 & 49.00 & 0.63 & 3.13 & 0.38 & 2.00 & 0.48 & 1.52 \\
        \bottomrule
        \end{tabular}
    }
    \vspace{-5pt}
\end{table}

We compare two embedding extraction strategies for LALMs. 
The first is the prompt-based last token approach described in Section~\ref{subsec:lalm}, where we append a summarization prompt to the input and extract the hidden state of the last token as the embedding. 
The second is mean pooling, where we directly average all hidden states across the sequence without using the summarization prompt.
Table~\ref{tab:pooling} presents the results.

The prompt-based last token approach generally outperforms mean pooling on DailyTalk and Synthetic, particularly for Qwen-series models. 
Qwen3-Omni exhibits the largest performance gap, with the last token embedding achieving substantially higher recall than mean pooling on semantic retrieval tasks. 
This suggests that the summarization prompt effectively guides the model to produce more discriminative representations for content-based retrieval. 
On VCTK and Expresso, both strategies yield similarly low performance, indicating that neither approach enables LALMs to effectively capture paralinguistic attributes such as speaker identity and speaking style.

These results demonstrate that the prompt-based last token strategy is more suitable for leveraging LALMs in retrieval tasks, though overall performance remains limited without instruction-aware retrieval training.

\subsection{Layer-wise Analysis of Self-Supervised Speech Models}
\label{subsec:ssl_layerwise}
\begin{figure}[htbp]
    \centering
    \vspace{-10pt}
    \begin{subfigure}{0.49\linewidth}
        \centering
        \includegraphics[width=\linewidth]{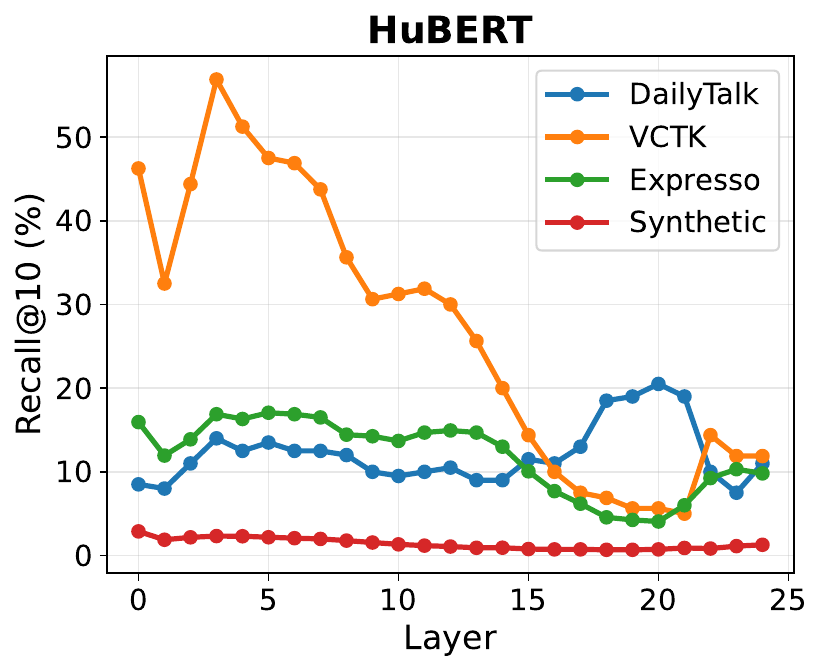}
    \end{subfigure}
    \begin{subfigure}{0.49\linewidth}
        \centering
        \includegraphics[width=\linewidth]{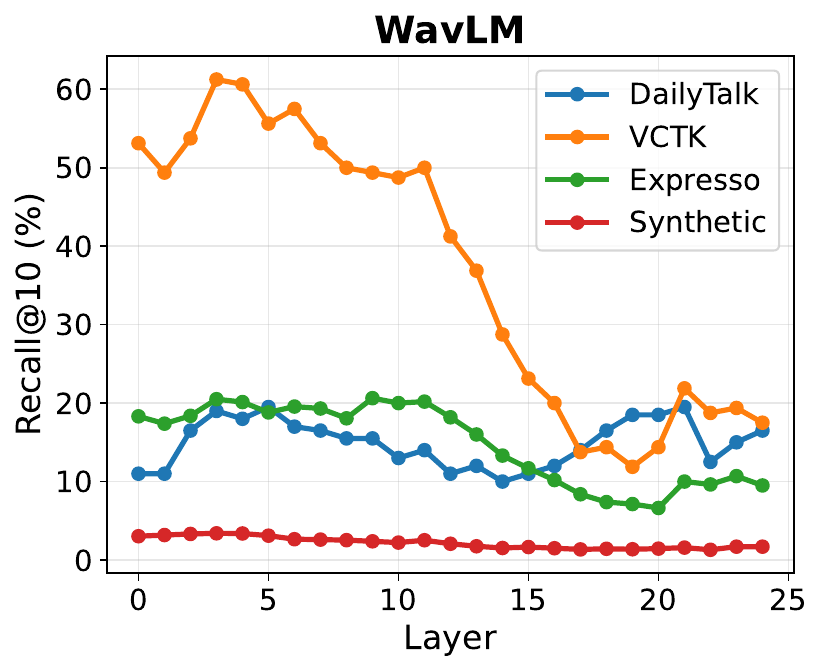}
    \end{subfigure}
    \vspace{-10pt}
    \caption{Layer-wise analysis of self-supervised speech models.}
    \label{fig:ssl_layerwise}
    \vspace{-5pt}
\end{figure}
To understand how different layers of self-supervised speech models encode information relevant to various retrieval tasks, we analyze layer-wise performance for HuBERT-Large and WavLM-Large. 
As described in Section~\ref{subsec:ssl}, we extract representations from different layers and compute retrieval performance. 
Figure~\ref{fig:ssl_layerwise} presents the results.

For both HuBERT and WavLM, performance on VCTK peaks in the lower to middle layers and then declines in deeper layers, indicating that speaker-related information is primarily encoded early in the network. 
DailyTalk shows moderate performance with peaks at specific layers, suggesting that semantic information emerges more clearly in intermediate to higher layers. 
Expresso follows a similar but lower trend, reflecting a balance between speaker and style representations across layers.

In contrast, Synthetic consistently exhibits the lowest recall across all layers for both models, highlighting the challenge of modeling complex instructions that combine semantics, speaker, style, and environmental factors. 
These findings align with prior work~\cite{pasad2021layer}, suggesting that different layers capture different types of information, and no single layer provides optimal representations for all retrieval intents in instruction-aware speech retrieval.

\subsection{Analysis for Contrastive Audio-Language Embeddings}
\label{subsec:clap_modality}
\begin{table}[hbtp]
    \centering
    \vspace{-10pt}
    \caption{Modality analysis for a contrastive audio-language model. $A$ and $T$ denote audio and text, respectively. $X \rightarrow Y$ indicates that the query is encoded with the $X$ encoder and the document with the $Y$ encoder.}
    \label{tab:clap}
    \vspace{-5pt}
    \resizebox{\linewidth}{!}{
        \begin{tabular}{lcccccccc}
        \toprule
        & \multicolumn{2}{c}{\textbf{DailyTalk}} & \multicolumn{2}{c}{\textbf{VCTK}} & \multicolumn{2}{c}{\textbf{Expresso}} & \multicolumn{2}{c}{\textbf{Synthetic}} \\
        \cmidrule(lr){2-3} \cmidrule(lr){4-5} \cmidrule(lr){6-7} \cmidrule(lr){8-9}
        \textbf{Modality} & \textbf{R@10} & \textbf{R@50} & \textbf{R@10} & \textbf{R@50} & \textbf{R@10} & \textbf{R@50} & \textbf{R@10} & \textbf{R@50} \\
        \midrule
        A$\rightarrow$A & 5.00 & 13.00 & \textbf{16.88} & \textbf{31.88} & \textbf{13.88} & \textbf{20.94} & \textbf{3.28} & \textbf{8.15} \\
        T$\rightarrow$A & 0.00 & 1.00 & 1.25 & 4.38 & 3.13 & 6.44 & 0.45 & 1.70 \\
        A$\rightarrow$T & 0.00 & 1.50 & 1.88 & 4.38 & 1.94 & 5.63 & 0.80 & 2.22 \\
        T$\rightarrow$T & \textbf{7.00} & \textbf{17.00} & 0.63 & 5.00 & 0.81 & 5.50 & 0.72 & 2.70 \\
        \bottomrule
        \end{tabular}
    }
    \vspace{-5pt}
\end{table}
To better understand the modality behavior of contrastive audio-language models, we analyze CLAP performance across different encoder configurations. 
As Section~\ref{subsec:clap} describes, contrastive audio-language model provides separate text and audio encoders that can be combined in various ways. 
We evaluate four configurations based on the encoder choice for the query and document sides.
For configurations that use the text encoder, we first convert spoken utterances to text via ASR and audio captioning.
When the instruction $z$ is available, we concatenate it with the text input for configurations that employ the text encoder on the query side.
For configurations that use the audio encoder on the query side, we do not include $z$ because the audio encoder does not accept textual input.
Table~\ref{tab:clap} summarizes the results.

Audio-to-audio retrieval outperforms other configurations on VCTK, Expresso, and Synthetic, indicating that the contrastive audio-language model effectively captures acoustic attributes such as speaker identity, speaking style, and environmental sounds within the audio modality.
Text-to-text retrieval achieves the best performance on DailyTalk, which primarily emphasizes semantic understanding rather than acoustic details.

In contrast, cross-modal retrieval consistently yields low recall across all datasets.
Both cross-modal configurations underperform their single-modality counterparts, highlighting the limited cross-modal alignment of the contrastive audio-language model for instruction-aware speech retrieval.

\section{Conclusion}
\label{sec:conclusion}
We introduced INSPIRE, the first benchmark for instruction-aware speech retrieval, formalizing a paradigm where natural-language instructions dynamically describe relevance across semantic, speaker, style, and environmental attributes. 
Our evaluation reveals a clear modality specialization gap: cascaded pipelines excel at semantic matching but struggle with paralinguistic attributes, while self-supervised speech models capture acoustic properties but lack instruction sensitivity. 
LALMs and pre-trained language model approaches do not bridge this gap. 
The most challenging multi-attribute scenarios remain largely unsolved by all methods. 
These findings motivate future work on unified multimodal embedding models with instruction-aware training that jointly preserve fine-grained acoustic information and support compositional instruction following.
\section{Generative AI Use Disclosure}
We use generative AI tools exclusively for polishing the manuscript.
We maintain full responsibility for the study’s design, data analysis, and scientific interpretations, which remain entirely independent of AI influence.
\section{Acknowledgements}
We thank Tzu-Han Lin, Yu-Xiang Lin, and Cheng-Han Chiang from National Taiwan University for their helpful discussions and feedback. 
This work was financially supported by the National Science and Technology Council (NSTC) in Taiwan. 
We thank to National Center for High-performance Computing (NCHC) of National Applied Research Laboratories (NARLabs) in Taiwan for providing computational and storage resources. 
This work was supported by the Ministry of Education (MOE) of Taiwan under the project Taiwan Centers of Excellence in Artificial Intelligence, through the NTU Artificial Intelligence Center of Research Excellence (NTU AI-CoRE).

\bibliographystyle{IEEEtran}
\bibliography{reference}

@inproceedings{ao2018query,
  title={Query-by-example spoken term detection using attention-based multi-hop networks},
  author={Ao, Chia-Wei and Lee, Hung-yi},
  booktitle={2018 IEEE International Conference on Acoustics, Speech and Signal Processing (ICASSP)},
  pages={6264--6268},
  year={2018},
  organization={IEEE}
}

@inproceedings{ram2018cnn,
  title={CNN Based Query by Example Spoken Term Detection.},
  author={Ram, Dhananjay and Miculicich, Lesly and Bourlard, Herv{\'e}},
  booktitle={Interspeech},
  pages={92--96},
  year={2018}
}

@article{ram2020neural,
  title={Neural network based end-to-end query by example spoken term detection},
  author={Ram, Dhananjay and Miculicich, Lesly and Bourlard, Herv{\'e}},
  journal={IEEE/ACM Transactions on Audio, Speech, and Language Processing},
  volume={28},
  pages={1416--1427},
  year={2020},
  publisher={IEEE}
}

@inproceedings{witbrock1997speech,
  title={Speech recognition and information retrieval: Experiments in retrieving spoken documents},
  author={Witbrock, Michael and Hauptmann, Alexander G},
  booktitle={Proceedings of the DARPA speech recognition workshop},
  volume={97},
  year={1997}
}

@article{garofolo2000trec,
  title={The TREC Spoken Document Retrieval Track: A Success Story.},
  author={Garofolo, John S and Auzanne, Cedric GP and Voorhees, Ellen M and others},
  journal={NIST SPECIAL PUBLICATION SP},
  volume={500},
  number={246},
  pages={107--130},
  year={2000},
  publisher={NATIONAL INSTIUTE OF STANDARDS \& TECHNOLOGY}
}

@article{chelba2008retrieval,
  title={Retrieval and browsing of spoken content},
  author={Chelba, Ciprian and Hazen, Timothy J and Saraclar, Murat},
  journal={IEEE Signal Processing Magazine},
  volume={25},
  number={3},
  pages={39--49},
  year={2008},
  publisher={IEEE}
}

@article{larson2012spoken,
  title={Spoken content retrieval: A survey of techniques and technologies},
  author={Larson, Martha and Jones, Gareth JF},
  journal={Foundations and Trends{\textregistered} in Information Retrieval},
  volume={5},
  number={4--5},
  pages={235--422},
  year={2012},
  publisher={Now Publishers, Inc.}
}

@article{lee2015spoken,
  title={Spoken content retrieval—beyond cascading speech recognition with text retrieval},
  author={Lee, Lin-shan and Glass, James and Lee, Hung-yi and Chan, Chun-an},
  journal={IEEE/ACM Transactions on Audio, Speech, and Language Processing},
  volume={23},
  number={9},
  pages={1389--1420},
  year={2015},
  publisher={IEEE}
}

@inproceedings{lin2024speechdpr,
  title={Speechdpr: End-to-end spoken passage retrieval for open-domain spoken question answering},
  author={Lin, Chyi-Jiunn and Lin, Guan-Ting and Chuang, Yung-Sung and Wu, Wei-Lun and Li, Shang-Wen and Mohamed, Abdelrahman and Lee, Hung-yi and Lee, Lin-Shan},
  booktitle={2024 IEEE International Conference on Acoustics, Speech and Signal Processing (ICASSP)},
  pages={12476--12480},
  year={2024},
  organization={IEEE}
}

@inproceedings{hu25b_interspeech,
  title     = {Vela: Scalable Embeddings with Voice Large Language Models for Multimodal Retrieval},
  author    = {Ruofan Hu and Yan Xia and Minjie Hong and Jieming Zhu and Bo Chen and Xiaoda Yang and Minghui Fang and Tao Jin},
  year      = {2025},
  booktitle = {Interspeech 2025},
  pages     = {2640--2644},
  doi       = {10.21437/Interspeech.2025-159},
  issn      = {2958-1796},
}

@inproceedings{turmo2009overview,
  title={Overview of QAST 2009},
  author={Turmo, Jordi and Comas, Pere R and Rosset, Sophie and Galibert, Olivier and Moreau, Nicolas and Mostefa, Djamel and Rosso, Paolo and Buscaldi, Davide},
  booktitle={Workshop of the Cross-Language Evaluation Forum for European Languages},
  pages={197--211},
  year={2009},
  organization={Springer}
}

@article{comas2012sibyl,
  title={Sibyl, a factoid question-answering system for spoken documents},
  author={Comas, Pere R and Turmo, Jordi and M{\`a}rquez, Llu{\'\i}s},
  journal={ACM Transactions on Information Systems (TOIS)},
  volume={30},
  number={3},
  pages={1--40},
  year={2012},
  publisher={ACM New York, NY, USA}
}

@article{li2018spoken,
  title={Spoken SQuAD: A study of mitigating the impact of speech recognition errors on listening comprehension},
  author={Li, Chia-Hsuan and Wu, Szu-Lin and Liu, Chi-Liang and Lee, Hung-yi},
  journal={arXiv preprint arXiv:1804.00320},
  year={2018}
}

@inproceedings{lee2018odsqa,
  title={ODSQA: Open-domain spoken question answering dataset},
  author={Lee, Chia-Hsuan and Wang, Shang-Ming and Chang, Huan-Cheng and Lee, Hung-Yi},
  booktitle={2018 IEEE Spoken Language Technology Workshop (SLT)},
  pages={949--956},
  year={2018},
  organization={IEEE}
}

@inproceedings{lin22c_interspeech,
  title     = {DUAL: Discrete Spoken Unit Adaptive Learning for Textless Spoken Question Answering},
  author    = {Guan-Ting Lin and Yung-Sung Chuang and Ho-Lam Chung and Shu-wen Yang and Hsuan-Jui Chen and Shuyan Annie Dong and Shang-Wen Li and Abdelrahman Mohamed and Hung-yi Lee and Lin-shan Lee},
  year      = {2022},
  booktitle = {Interspeech 2022},
  pages     = {5165--5169},
  doi       = {10.21437/Interspeech.2022-612},
  issn      = {2958-1796},
}

@inproceedings{you2022end,
  title={End-to-end Spoken Conversational Question Answering: Task, Dataset and Model},
  author={You, Chenyu and Chen, Nuo and Liu, Fenglin and Ge, Shen and Wu, Xian and Zou, Yuexian},
  booktitle={Findings of the Association for Computational Linguistics: NAACL 2022},
  pages={1219--1232},
  year={2022}
}

@inproceedings{shon2023slue,
  title={SLUE phase-2: A benchmark suite of diverse spoken language understanding tasks},
  author={Shon, Suwon and Arora, Siddhant and Lin, Chyi-Jiunn and Pasad, Ankita and Wu, Felix and Sharma, Roshan and Wu, Wei-Lun and Lee, Hung-yi and Livescu, Karen and Watanabe, Shinji},
  booktitle={Proceedings of the 61st Annual Meeting of the Association for Computational Linguistics (Volume 1: Long Papers)},
  pages={8906--8937},
  year={2023}
}

@inproceedings{asai2023task,
  title={Task-aware retrieval with instructions},
  author={Asai, Akari and Schick, Timo and Lewis, Patrick and Chen, Xilun and Izacard, Gautier and Riedel, Sebastian and Hajishirzi, Hannaneh and Yih, Wen-tau},
  booktitle={Findings of the Association for Computational Linguistics: ACL 2023},
  pages={3650--3675},
  year={2023}
}

@inproceedings{su2023one,
  title={One embedder, any task: Instruction-finetuned text embeddings},
  author={Su, Hongjin and Shi, Weijia and Kasai, Jungo and Wang, Yizhong and Hu, Yushi and Ostendorf, Mari and Yih, Wen-tau and Smith, Noah A and Zettlemoyer, Luke and Yu, Tao},
  booktitle={Findings of the Association for Computational Linguistics: ACL 2023},
  pages={1102--1121},
  year={2023}
}

@inproceedings{lee2025nvembed,
    title={{NV}-Embed: Improved Techniques for Training {LLM}s as Generalist Embedding Models},
    author={Chankyu Lee and Rajarshi Roy and Mengyao Xu and Jonathan Raiman and Mohammad Shoeybi and Bryan Catanzaro and Wei Ping},
    booktitle={The Thirteenth International Conference on Learning Representations},
    year={2025},
}

@article{lee2025gemini,
  title={Gemini embedding: Generalizable embeddings from gemini},
  author={Lee, Jinhyuk and Chen, Feiyang and Dua, Sahil and others},
  journal={arXiv preprint arXiv:2503.07891},
  year={2025}
}

@inproceedings{wu2021fashion,
  title={Fashion iq: A new dataset towards retrieving images by natural language feedback},
  author={Wu, Hui and Gao, Yupeng and Guo, Xiaoxiao and Al-Halah, Ziad and Rennie, Steven and Grauman, Kristen and Feris, Rogerio},
  booktitle={Proceedings of the IEEE/CVF Conference on computer vision and pattern recognition},
  pages={11307--11317},
  year={2021}
}

@inproceedings{liu2021image,
  title={Image retrieval on real-life images with pre-trained vision-and-language models},
  author={Liu, Zheyuan and Rodriguez-Opazo, Cristian and Teney, Damien and Gould, Stephen},
  booktitle={Proceedings of the IEEE/CVF international conference on computer vision},
  pages={2125--2134},
  year={2021}
}

@inproceedings{baldrati2023zero,
  title={Zero-shot composed image retrieval with textual inversion},
  author={Baldrati, Alberto and Agnolucci, Lorenzo and Bertini, Marco and Del Bimbo, Alberto},
  booktitle={Proceedings of the IEEE/CVF International Conference on Computer Vision},
  pages={15338--15347},
  year={2023}
}

@inproceedings{saito2023pic2word,
  title={Pic2word: Mapping pictures to words for zero-shot composed image retrieval},
  author={Saito, Kuniaki and Sohn, Kihyuk and Zhang, Xiang and Li, Chun-Liang and Lee, Chen-Yu and Saenko, Kate and Pfister, Tomas},
  booktitle={Proceedings of the IEEE/CVF Conference on Computer Vision and Pattern Recognition},
  pages={19305--19314},
  year={2023}
}

@inproceedings{zhang2024magiclens,
    title={MagicLens: Self-Supervised Image Retrieval with Open-Ended Instructions},
    author={Kai Zhang and Yi Luan and Hexiang Hu and Kenton Lee and Siyuan Qiao and Wenhu Chen and Yu Su and Ming-Wei Chang},
    booktitle={Forty-first International Conference on Machine Learning},
    year={2024},
}

@article{jiang2024e5,
  title={E5-v: Universal embeddings with multimodal large language models},
  author={Jiang, Ting and Song, Minghui and Zhang, Zihan and Huang, Haizhen and Deng, Weiwei and Sun, Feng and Zhang, Qi and Wang, Deqing and Zhuang, Fuzhen},
  journal={arXiv preprint arXiv:2407.12580},
  year={2024}
}

@inproceedings{zhang2025bridging,
  title={Bridging Modalities: Improving Universal Multimodal Retrieval by Multimodal Large Language Models},
  author={Zhang, Xin and Zhang, Yanzhao and Xie, Wen and Li, Mingxin and Dai, Ziqi and Long, Dingkun and Xie, Pengjun and Zhang, Meishan and Li, Wenjie and Zhang, Min},
  booktitle={Proceedings of the Computer Vision and Pattern Recognition Conference},
  pages={9274--9285},
  year={2025}
}

@inproceedings{zhou2025megapairs,
  title={Megapairs: Massive data synthesis for universal multimodal retrieval},
  author={Zhou, Junjie and Xiong, Yongping and Liu, Zheng and Liu, Ze and Xiao, Shitao and Wang, Yueze and Zhao, Bo and Zhang, Chen Jason and Lian, Defu},
  booktitle={Proceedings of the 63rd Annual Meeting of the Association for Computational Linguistics (Volume 1: Long Papers)},
  pages={19076--19095},
  year={2025}
}

@inproceedings{jiang2025vlmvec,
    title={{VLM}2Vec: Training Vision-Language Models for Massive Multimodal Embedding Tasks},
    author={Ziyan Jiang and Rui Meng and Xinyi Yang and Semih Yavuz and Yingbo Zhou and Wenhu Chen},
    booktitle={The Thirteenth International Conference on Learning Representations},
    year={2025},
}

@article{meng2025vlm2vec,
  title={Vlm2vec-v2: Advancing multimodal embedding for videos, images, and visual documents},
  author={Meng, Rui and Jiang, Ziyan and Liu, Ye and others},
  journal={arXiv preprint arXiv:2507.04590},
  year={2025}
}

@article{li2026qwen3,
  title={Qwen3-VL-Embedding and Qwen3-VL-Reranker: A Unified Framework for State-of-the-Art Multimodal Retrieval and Ranking},
  author={Li, Mingxin and Zhang, Yanzhao and Long, Dingkun and others},
  journal={arXiv preprint arXiv:2601.04720},
  year={2026}
}

@inproceedings{shor20_interspeech,
  title     = {Towards Learning a Universal Non-Semantic Representation of Speech},
  author    = {Joel Shor and Aren Jansen and Ronnie Maor and Oran Lang and Omry Tuval and Félix de Chaumont Quitry and Marco Tagliasacchi and Ira Shavitt and Dotan Emanuel and Yinnon Haviv},
  year      = {2020},
  booktitle = {Interspeech 2020},
  pages     = {140--144},
  doi       = {10.21437/Interspeech.2020-1242},
  issn      = {2958-1796},
}

@inproceedings{yang21c_interspeech,
  title     = {SUPERB: Speech Processing Universal PERformance Benchmark},
  author    = {Shu-wen Yang and Po-Han Chi and Yung-Sung Chuang and others},
  year      = {2021},
  booktitle = {Interspeech 2021},
  pages     = {1194--1198},
  doi       = {10.21437/Interspeech.2021-1775},
  issn      = {2958-1796},
}

@inproceedings{wang2022towards,
  title={Towards learning universal audio representations},
  author={Wang, Luyu and Luc, Pauline and Wu, Yan and others},
  booktitle={2022 IEEE International Conference on Acoustics, Speech and Signal Processing (ICASSP)},
  pages={4593--4597},
  year={2022},
  organization={IEEE}
}

@inproceedings{
    heigold2025massive,
    title={Massive Sound Embedding Benchmark ({MSEB})},
    author={Georg Heigold and Ehsan Variani and Tom Bagby and Cyril Allauzen and Ji Ma and Shankar Kumar and Michael Riley},
    booktitle={The Thirty-ninth Annual Conference on Neural Information Processing Systems Datasets and Benchmarks Track},
    year={2025},
}

@article{assadi2026maeb,
  title={MAEB: Massive Audio Embedding Benchmark},
  author={Assadi, Adnan El and Chung, Isaac and Xiao, Chenghao and others},
  journal={arXiv preprint arXiv:2602.16008},
  year={2026}
}

@inproceedings{lee2023dailytalk,
  title={Dailytalk: Spoken dialogue dataset for conversational text-to-speech},
  author={Lee, Keon and Park, Kyumin and Kim, Daeyoung},
  booktitle={2023 IEEE International Conference on Acoustics, Speech and Signal Processing (ICASSP)},
  pages={1--5},
  year={2023},
  organization={IEEE}
}

@misc{yamagishi2019vctk,
  author={Yamagishi, Junichi and Veaux, Christophe and MacDonald, Kirsten},
  title={ {CSTR VCTK Corpus}: English Multi-speaker Corpus for {CSTR} Voice Cloning Toolkit (version 0.92)},
  publisher={University of Edinburgh. The Centre for Speech Technology Research (CSTR)},
  year=2019,
  doi={10.7488/ds/2645},
}

@inproceedings{nguyen23_interspeech,
  title     = {Expresso: A Benchmark and Analysis of Discrete Expressive Speech Resynthesis},
  author    = {Tu Anh Nguyen and Wei-Ning Hsu and Antony D'Avirro and Bowen Shi and Itai Gat and Maryam Fazel-Zarani and Tal Remez and Jade Copet and Gabriel Synnaeve and Michael Hassid and Felix Kreuk and Yossi Adi and Emmanuel Dupoux},
  year      = {2023},
  booktitle = {Interspeech 2023},
  pages     = {4823--4827},
  doi       = {10.21437/Interspeech.2023-1905},
  issn      = {2958-1796},
}

@article{kwiatkowski2019natural,
  title={Natural questions: a benchmark for question answering research},
  author={Kwiatkowski, Tom and Palomaki, Jennimaria and Redfield, Olivia  and others},
  journal={Transactions of the Association for Computational Linguistics},
  volume={7},
  pages={453--466},
  year={2019},
  publisher={MIT Press One Rogers Street, Cambridge, MA 02142-1209, USA journals-info~…}
}

@misc{openai2024gpt4ocard,
  title={GPT-4o System Card}, 
  author={OpenAI},
  year={2024},
  eprint={2410.21276},
  archivePrefix={arXiv},
  primaryClass={cs.CL},
}

@inproceedings{piczak2015esc,
  title={ESC: Dataset for environmental sound classification},
  author={Piczak, Karol J},
  booktitle={Proceedings of the 23rd ACM international conference on Multimedia},
  pages={1015--1018},
  year={2015}
}

@misc{singh2025openaigpt5card,
      title={OpenAI GPT-5 System Card}, 
      author={OpenAI},
      year={2025},
      eprint={2601.03267},
      archivePrefix={arXiv},
      primaryClass={cs.CL},
}

@inproceedings{radford2023robust,
  title={Robust speech recognition via large-scale weak supervision},
  author={Radford, Alec and Kim, Jong Wook and Xu, Tao and Brockman, Greg and McLeavey, Christine and Sutskever, Ilya},
  booktitle={International conference on machine learning},
  pages={28492--28518},
  year={2023},
  organization={PMLR}
}

@inproceedings{desplanques2020ecapa,
  title={ECAPA-TDNN: Emphasized Channel Attention, propagation and aggregation in TDNN based speaker verification},
  author={Desplanques, Brecht and Thienpondt, Jenthe and Demuynck, Kris},
  booktitle={Interspeech 2020},
  pages={3830--3834},
  year={2020}
}

@article{ravanelli2021speechbrain,
  title={SpeechBrain: A general-purpose speech toolkit},
  author={Ravanelli, Mirco and Parcollet, Titouan and Plantinga, Peter and others},
  journal={arXiv preprint arXiv:2106.04624},
  year={2021}
}

@article{platt1999probabilistic,
  title={Probabilistic outputs for support vector machines and comparisons to regularized likelihood methods},
  author={Platt, John and others},
  journal={Advances in large margin classifiers},
  volume={10},
  number={3},
  pages={61--74},
  year={1999},
  publisher={Cambridge, MA}
}

@article{chang2011libsvm,
  title={LIBSVM: A library for support vector machines},
  author={Chang, Chih-Chung and Lin, Chih-Jen},
  journal={ACM transactions on intelligent systems and technology (TIST)},
  volume={2},
  number={3},
  pages={1--27},
  year={2011},
  publisher={Acm New York, NY, USA}
}

@inproceedings{ma2024emotion2vec,
  title={emotion2vec: Self-supervised pre-training for speech emotion representation},
  author={Ma, Ziyang and Zheng, Zhisheng and Ye, Jiaxin and Li, Jinchao and Gao, Zhifu and Zhang, Shiliang and Chen, Xie},
  booktitle={Findings of the Association for Computational Linguistics: ACL 2024},
  pages={15747--15760},
  year={2024}
}

@inproceedings{saeki22c_interspeech,
  title     = {UTMOS: UTokyo-SaruLab System for VoiceMOS Challenge 2022},
  author    = {Takaaki Saeki and Detai Xin and Wataru Nakata and Tomoki Koriyama and Shinnosuke Takamichi and Hiroshi Saruwatari},
  year      = {2022},
  booktitle = {Interspeech 2022},
  pages     = {4521--4525},
  doi       = {10.21437/Interspeech.2022-439},
  issn      = {2958-1796},
}

@article{xu2025qwen2,
  title={Qwen2. 5-omni technical report},
  author={Xu, Jin and Guo, Zhifang and He, Jinzheng and others},
  journal={arXiv preprint arXiv:2503.20215},
  year={2025}
}

@article{xu2025qwen3,
  title={Qwen3-Omni Technical Report},
  author={Jin Xu and Zhifang Guo and Hangrui Hu and others},
  journal={arXiv preprint arXiv:2509.17765},
  year={2025}
}

@article{liu2025voxtral,
  title={Voxtral},
  author={Liu, Alexander H and Ehrenberg, Andy and Lo, Andy and others},
  journal={arXiv preprint arXiv:2507.13264},
  year={2025}
}

@article{goel2025audio,
  title={Audio flamingo 3: Advancing audio intelligence with fully open large audio language models},
  author={Goel, Arushi and Ghosh, Sreyan and Kim, Jaehyeon and others},
  journal={arXiv preprint arXiv:2507.08128},
  year={2025}
}

@article{robertson2009probabilistic,
  title={The probabilistic relevance framework: BM25 and beyond},
  author={Robertson, Stephen and Zaragoza, Hugo and others},
  journal={Foundations and trends{\textregistered} in information retrieval},
  volume={3},
  number={4},
  pages={333--389},
  year={2009},
  publisher={Now Publishers, Inc.}
}

@inproceedings{reimers2019sentence,
  title={Sentence-BERT: Sentence Embeddings using Siamese BERT-Networks},
  author={Reimers, Nils and Gurevych, Iryna},
  booktitle={Proceedings of the 2019 Conference on Empirical Methods in Natural Language Processing and the 9th International Joint Conference on Natural Language Processing (EMNLP-IJCNLP)},
  pages={3982--3992},
  year={2019}
}

@inproceedings{wang2024improving,
  title={Improving text embeddings with large language models},
  author={Wang, Liang and Yang, Nan and Huang, Xiaolong and Yang, Linjun and Majumder, Rangan and Wei, Furu},
  booktitle={Proceedings of the 62nd Annual Meeting of the Association for Computational Linguistics (Volume 1: Long Papers)},
  pages={11897--11916},
  year={2024}
}

@article{zhang2025qwen3,
  title={Qwen3 Embedding: Advancing Text Embedding and Reranking Through Foundation Models},
  author={Zhang, Yanzhao and Li, Mingxin and Long, Dingkun and others},
  journal={arXiv preprint arXiv:2506.05176},
  year={2025}
}

@article{hsu2021hubert,
  title={Hubert: Self-supervised speech representation learning by masked prediction of hidden units},
  author={Hsu, Wei-Ning and Bolte, Benjamin and Tsai, Yao-Hung Hubert and Lakhotia, Kushal and Salakhutdinov, Ruslan and Mohamed, Abdelrahman},
  journal={IEEE/ACM transactions on audio, speech, and language processing},
  volume={29},
  pages={3451--3460},
  year={2021},
  publisher={IEEE}
}

@article{chen2022wavlm,
  title={Wavlm: Large-scale self-supervised pre-training for full stack speech processing},
  author={Chen, Sanyuan and Wang, Chengyi and Chen, Zhengyang and others},
  journal={IEEE Journal of Selected Topics in Signal Processing},
  volume={16},
  number={6},
  pages={1505--1518},
  year={2022},
  publisher={IEEE}
}

@inproceedings{wu2023large,
  title={Large-scale contrastive language-audio pretraining with feature fusion and keyword-to-caption augmentation},
  author={Wu, Yusong and Chen, Ke and Zhang, Tianyu and Hui, Yuchen and Berg-Kirkpatrick, Taylor and Dubnov, Shlomo},
  booktitle={2023 IEEE International Conference on Acoustics, Speech and Signal Processing (ICASSP)},
  pages={1--5},
  year={2023},
  organization={IEEE}
}

@misc{openaiemb,
  key     = {OpenAI},
  title   = {New embedding models and API updates},
  year    = 2024,
  url     = {https://openai.com/index/new-embedding-models-and-api-updates/},
}

@misc{gemini-3-pro,
	title = {A new era of intelligence with Gemini 3},
	author = {Google},
	year = {2025},
	url = {https://blog.google/products/gemini/gemini-3}
}

@inproceedings{jiang2024scaling,
  title={Scaling sentence embeddings with large language models},
  author={Jiang, Ting and Huang, Shaohan and Luan, Zhongzhi and Wang, Deqing and Zhuang, Fuzhen},
  booktitle={Findings of the association for computational linguistics: EMNLP 2024},
  pages={3182--3196},
  year={2024}
}

@inproceedings{elizalde2023clap,
  title={Clap learning audio concepts from natural language supervision},
  author={Elizalde, Benjamin and Deshmukh, Soham and Al Ismail, Mahmoud and Wang, Huaming},
  booktitle={2023 IEEE International Conference on Acoustics, Speech and Signal Processing (ICASSP)},
  pages={1--5},
  year={2023},
  organization={IEEE}
}

@inproceedings{pasad2021layer,
  title={Layer-wise analysis of a self-supervised speech representation model},
  author={Pasad, Ankita and Chou, Ju-Chieh and Livescu, Karen},
  booktitle={2021 IEEE Automatic Speech Recognition and Understanding Workshop (ASRU)},
  pages={914--921},
  year={2021},
  organization={IEEE}
}

\end{document}